\documentclass{pasj01}
\draft 
\Received{$\langle$2020 March 19$\rangle$}
\Accepted{$\langle$2020 July 3$\rangle$}
\Published{$\langle$publication date$\rangle$}

\newcommand{\tabref}[1]{Table~\ref{#1}}

\newcommand{\figref}[1]{Figure~\ref{#1}}

\usepackage[hang,small,bf]{caption}
\usepackage[subrefformat=parens]{subcaption}
\begin{document}

\title{Measurements of the Ca\,\emissiontype{II} infrared triplet emission lines of pre-main-sequence stars}
\author{${}^1$Mai Yamashita, ${}^1$Yoichi Itoh, ${}^2$Yuhei Takagi}%
\altaffiltext{}{${}^1$Nishi-Harima Astronomical Observatory, Center for Astronomy, University of Hyogo, 407-2 Nishigaichi, Sayo, Sayo, Hyogo 679-5313 }
\altaffiltext{}{${}^2$Subaru Telescope, National Astoronomical Observatory of Japan, 650 North A'ohoku Place, Hilo, HI 96720, U.S.A.}
\email{yamashita@nhao.jp}


\maketitle

\begin{abstract}
We investigated the chromospheric activity of 60 pre-main-sequence (PMS) stars in four molecular clouds and five moving groups. It is considered that strong chromospheric activity is driven by the dynamo processes generated by the stellar rotation. In contrast, several researchers have pointed out that the chromospheres of PMS stars are activated by mass accretion from their protoplanetary disks. In this study, the Ca\,\emissiontype{II} infrared triplet (IRT) emission lines were investigated utilizing medium- and high-resolution spectroscopy. The observations were conducted with Nayuta/MALLS and Subaru/HDS. Additionally, archive data obtained by Keck/HIRES, VLT/UVES, and VLT/X-Shooter was used. The small ratios of the equivalent widths indicate that Ca\,\emissiontype{II} IRT emission lines arise primarily in dense chromospheric regions. Seven PMS stars show broad emission lines. Among them, four PMS stars have more than one order of magnitude brighter emission line fluxes compared to the low-mass stars in young open clusters. The four PMS stars have a high mass accretion rate, which indicates that the broad and strong emission results from a large mass accretion. However, most PMS stars exhibit narrow emission lines. No significant correlation was found between the accretion rate and flux of the emission line. The ratios of the surface flux of the Ca\,\emissiontype{II} IRT lines to the stellar bolometric luminosity, $R^{\prime}_{\rm IRT}$, of the PMS stars with narrow emission lines are as large as the largest $R^{\prime}_{\rm IRT}$ of the low-mass stars in the young open clusters. This result indicates that most PMS stars, even in the classical T Tauri star stage, have chromospheric activity similar to zero-age main-sequence stars. 
\end{abstract}

\chapter{Introduction}
\label{intro}

Chromosphere is the region between the photosphere and corona. The temperature of the chromosphere gradually increases with radial distance to the photosphere, in case of the Sun, it is from almost $4000 \, \mathrm{K}$ at the bottom to $8000 \, \mathrm{K}$ at the top (\cite{v81}). 
Because of these high temperature, atoms emit some permitted lines like H$\rm \alpha$ and Ca\,\emissiontype{II}. Several high spatial-resolution observations of the Sun show that in the chromosphere the above mentioned emission lines form under the influence of strong magnetic fields, such as faculae and plages near dark spots (\cite{l17}). 
The Hinode satellite has obtained space- and time-resolved images of the solar chromosphere and revealed its energetic activity (\cite{ka07}). Over a 30 year period, \citet{l07} showed that fluxes of the Ca\,\emissiontype{II} K, H$\rm \alpha$, and He\,\emissiontype{I} lines were strong when the solar magnetic activity was high. These observations revealed the relationship between the emission line strengths and the chromospheric activity induced by the magnetic field.

For main-sequence stars, chromospheric activity is often discussed in relation to stellar rotation.
It is claimed that chromospheric activity is driven by the magnetic field, which is generated by the dynamo process. \citet{w78} showed that chromospheric activity analogous to that associated with the  solar magnetic activity cycle is ubiquitous in stars along the lower main-sequence stars. As mentioned in \citet{noyes}, rotation is not the only parameter related to the dynamo process, the stellar mass, spectral type, depth of convection zone, and convective turnover time ($\tau_{\rm c}$) are also related to the process. 
These five parameters are included in the Rossby number ($N_{\rm R}$), which is defined as $P_{\rm obs}/\tau_{\rm c}$, where $P_{\rm obs}$ is the stellar rotational period.
According to standard dynamo theory in \citet{parker}, the magnetic activity at the stellar surface is getting stronger with increasing rotation speed and strength of differential rotation.

It is well established that a young stars are fast rotators (e.g. \cite{bouvier90}). \citet{s72} found that the luminosity of the Ca\,\emissiontype{II} emission lines and the rotational velocity of solar-mass stars with an age of $10^7 \, \mathrm{yr}$ are one order of magnitude larger than those of stars with $10^{10} \, \mathrm{yr}$. \citet{so93} revealed strong emission lines of the Ca\,\emissiontype{II} infrared triplet (IRT ; $\lambda 8498, 8542, 8662 \, \mathrm{\AA}$) of low-mass stars in a young open cluster, M45 (age $130 \pm 20 \, \mathrm{Myr}, 0.6-1.4 \, \mathrm{M_{\odot}}$). \citet{marsden09} observed low-mass stars in young open clusters, IC 2391 and IC 2602, including rapidly rotationing stars. The age of IC 2391 and IC 2602 is $50 \pm 5 \, \mathrm{Myr}$ and $30 \pm 5 \, \mathrm{Myr}$, respectively (\cite{st97} ; \cite{na04}). The cluster members are considered to be on the zero-age main-sequence (ZAMS) or in the last phase of evolving to ZAMS with a mass between $0.8 \, \mathrm{M_{\odot}}$ and $1.5 \, \mathrm{M_{\odot}}$. \citet{so93} and \citet{marsden09} calculated $R^{\prime}_{\rm IRT}$ from the equivalent widths (EQWs). $R^{\prime}_{\rm IRT}$ describes the ratio of the surface flux of the Ca\,\emissiontype{II} IRT emission lines to the stellar bolometric luminosity. 
They found that $R^{\prime}_{\rm IRT}$ is constant at levels of around $\log R^{\prime}_{\rm \lambda 8498} \sim -4.4$, $\log R^{\prime}_{\rm \lambda 8542} \sim -4.2$, and $\log R^{\prime}_{\rm \lambda 8662} \sim -4.3$ for stars with $\log N_{\rm R} \leq -1.1$.
These regions are called the saturated regime. For stars with $\log N_{\rm R} \geq -1.1$, $R^{\prime}_{\rm IRT}$ decreases with increasing $N_{\rm R}$. This region is called the unsaturated regime. \citet{marsden09} suggested that the chromosphere is completely filled by the emitting region for stars in the saturated regime.

The idea that the activity of the chromospheres is driven by the dynamo process is widely accepted not only for low-mass main-sequence stars but also for pre-main-sequence (PMS) stars. However several previous studies showed that a certain portion of T Tauri stars (TTSs) are slow rotators. \citet{kuhi} found that low-mass PMS stars ($\leq 1.5 \, \mathrm{M_{\odot}}$) generally rotate at less than $25 \, \mathrm{km \cdot s^{-1}}$. 
\citet{wb03} revealed that all of the low-mass TTSs in the Taurus-Auriga star forming region are slowly rotating ($v \sin i < 30 \, \mathrm{km \cdot s^{-1}}$). \citet{h86} confirmed and extended the findings of \citet{kuhi}; they found that about $30 \%$ of the $0.5 - 1.0 \, \mathrm{M_{\odot}}$ TTSs in Taurus have rotational velocities at or below $10 \, \mathrm{km \cdot s^{-1}}$, and half have $v \sin i$ values between $10$ and $15 \, \mathrm{km \cdot s^{-1}}$. However, previous studies also showed that a fraction of TTSs exhibit strong Ca\,\emissiontype{II} emission lines (\cite{h87}). \citet{bb93} constructed models of photosphere and chromosphere to fit the observed profiles of the Ca\,\emissiontype{II} IRT emission line for TTSs with low accretion rates but null veiling. Those deep chromosphere models explain the emission characteristics of TTSs, showing a narrow emission profile over a broad absorption feature (\cite{ca84}; \cite{bb96}). \citet{cg98} found that shock heating of the photosphere by emission of chromosphere result in a temperature inversion and the temperature increases in the chromosphere, as shown by deep chromospheric models.  In contrast, \citet{m12} found no clear correlation between the EQWs of the Ca\,\emissiontype{II} IRT emission lines and stellar rotation velocity. As mentioned above, the rotation velocity is not the only parameter expected to be related to the dynamo process. The effective temperature ($T_{\rm eff}$), $\tau_{\rm c}$, and the spectral distribution of the continuum flux change as PMS stars evolve. 

Another indicator for chromospheric activity of PMS stars is mass accretion from their protoplanetary disks. \citet{m12} found that the EQWs of the Ca\,\emissiontype{II} IRT emission lines decrease with stellar evolution from classical TTSs (CTTSs), transitional disk objects,  weak-line TTSs (WTTSs), to ZAMS stars. They also revealed that PMS stars with high mass accretion rates have strong Ca\,\emissiontype{II} emission lines.
\citet{mo05} investigated the chromospheric activity of CTTSs, very low-mass young stars ($0.075 \leq M_* <  0.15 \, \mathrm{M_{\odot}}$), and young brown dwarfs ($M_* \leq 0.075 \, \mathrm{M_{\odot}}$). The surface flux of the Ca\,\emissiontype{II} emission line at $\lambda 8662 \, \mathrm{\AA}$, $F^{\prime}_{\lambda 8662}$, exhibit correlation with the associated mass accretion rate, $\dot{M}$, for approximately 4 orders of magnitude. Hence, \citet{mo05} claimed that the Ca\,\emissiontype{II} emission line is an excellent quantitative measure for the accretion rate. 
\citet{hp92} carried out optical spectroscopy for 53 TTSs and 32 Herbig Ae/Be stars. They interpreted narrow emission lines such as Ca\,\emissiontype{II} and Mg\,\emissiontype{I} generated in the stellar chromosphere. The Ca\,\emissiontype{II} IRT emission lines of several TTSs and Herbig Ae/Be stars also have a broad line component. This profile can be well explained by the magnetospheric accretion model (e.g. \cite{mu98}).

In this study, we investigate the Ca\,\emissiontype{II} IRT emission lines of 60 PMS stars with medium- and high-resolution spectral data.
We compare the rotation-activity relationship of the PMS stars with that of low-mass stars in young open clusters.
The chromospheric activity of the cluster members is considered to be induced by dynamo activity.
In the next section, we describe the observation and the data reduction procedures.
In Section \ref{result}, we present the results, and in Section \ref{discussion}, we discuss the origin of the Ca\,\emissiontype{II} IRT emission lines and the emitting region on the stellar surface.

\chapter{Observations and Data Reduction}
\label{observ}

\section{Stellar parameters}

All targets investigated in this study are listed in \tabref{tab:objects1}. 
These 60 objects are associated with four molecular clouds or five moving groups; the Taurus-Auriga molecular cloud, the Orionis OB 1c association, the Upper Scorpius association, the Perseus molecular cloud, the TW Hydrae association, the $\eta$ Chamaeleontis cluster, the ''Cha-Near'' region, the $\beta$ Pictoris moving group, and the AB Doradus moving group. 
Hereafter, objects belonging to both molecular clouds and moving groups are called ''low-mass PMS stars''.
We did not observe binaries or triplets listed in \citet{ghez93}, \citet{k09}, \citet{k12}, \citet{ne95}, \citet{l93}, \citet{wa10} and \citet{zs04}.


{\scriptsize
\renewcommand{\tabcolsep}{4pt}  
\begin{longtable}{lccrcccccccccc} 
\caption{Physical parameters of the PMS stars.}
\label{tab:objects1}    
\hline\noalign{\vskip3pt} 
Object Name &   $i$  &  $(B-V)_0$ &   $A_V$ &  $L/L_{\rm \odot}$ &  $T_{\rm eff}$ &  dist &  $v \sin i$ &  $\tau_{\rm c} $ &  $\log \dot{M}$ &  $M_*$ &  $M_{\rm conv}$ &  Age & Tel. \\
{}  &  $\mathrm{mag}$ &  $\mathrm{mag}$  &  $\mathrm{mag}$    &   &   $\mathrm{K}$ &  $\mathrm{pc}$ & $\mathrm{km \cdot s^{-1}}$ & $\mathrm{\times 10^6 s}$ & $\mathrm{M_{\odot} \cdot yr^{-1}}$ & $\mathrm{M_{\odot}}$  &  $\mathrm{M_{\odot}}$ &  $\mathrm{Myr}$ & {} \\ 
(1)  &  (2) &  (3) &  (4) &  (5) &  (6) & (7) &  (8) &  (9) & (10) &  (11) &  (12) & (13) & (14)\\    [2pt] 
\hline\noalign{\vskip3pt} 
\endhead
\endfoot
\multicolumn{2}{@{}l@{}}{\hbox to0pt{\parbox{170mm}{\footnotesize 
{(1) LRL 72 : \citet{l16} ; } 
{(2) $i$- mag : UCAC4 Catalogue (\cite{za13}) ; } 
{(3) $B-V$ : \citet{v07}, \citet{ma07}, \citet{ma06}, \citet{he16}, \citet{hb88}, }  {\citet{ha03}, and \citet{d14} ; } 
{(4)$A_V$ : \citet{k09}, \citet{ma06}, and \citet{wa10}, } 
{(5)(6) Luminosity and $T_{\rm eff}$ : {\it Gaia} DR2 (\cite{g18}), \citet{palla}, \citet{kh95}, } 
 {and \citet{pe13} ;} 
{(7) Distance : {\it Gaia} DR2 (\cite{ba18}) ;} 
{(8) $v \sin i$ : \citet{g05}, \citet{n12}, \citet{to06}, and \citet{me11} ; } 
{(10) $\dot{M}$ : \citet{najita}, \citet{gu98}, \citet{wg02}, \citet{h98}, }  {Lawson et al. (2004), \citet{ca04}, and \citet{i13}} 
{(14) N : Nayuta Telescope ; S : Subaru Telescope ; K : Keck Telescope ; V : VLT} 
{$\rm (*)$ For RY Tau, $A_V$, $L/L_{\rm \odot}$, $T_{\rm eff}$, and the distance were taken from \citet{gar}.} 
}\hss}} 
\endlastfoot
\multicolumn{6}{l}{Taurus-Auriga molecular cloud} & & & & & & & & \\ \hline
AA Tau                  &   12.7  &   0.82 &   0.74 &               0.74 &           4060 &        136 &    12.7 &                             20 &           -8.48 &    0.5 &            0.0 &        1.0 & V \\ 
BP Tau                  &   11.0  &   1.01 &   0.34 &               0.46 &           4320 &        128 &    10.9 &                             14 &           -7.54 &    0.9 &            0.2 &        5.0  & V \\
CX Tau                  &   11.9  &   1.31 &   0.80 &               0.28 &           3417 &        127 &    19.1 &                             19 &           -8.97 &    0.3 &            0.0 &        1.0  & K \\
CoKu Tau4               &   12.6  &   1.30 &   1.75 &               0.11 &           4070 &        170 &    25.8 &                             11 &          -10.0 &    0.6 &            0.2 &       20.0  & K \\
DF Tau                  &   10.4  &   0.75 &   0.45 &               0.93 &           3665 &        124 &    18.4 &                             19 &           -7.62 &    0.3 &            0.0 &        0.3  & V \\
DG Tau                  &   11.5  &   0.54 &   1.30 &               0.34 &           3731 &        120 &    21.7 &                             19 &           -6.30 &    0.4 &            0.0 &        1.0  & N \\
DL Tau                  &   11.6  &   0.53 &   2.00 &               0.44 &           3998 &        158 &    19.0 &                             21 &           -6.79 &    0.6 &            0.0 &        2.0  & N \\
DM Tau                  &   12.6  &   0.82 &   1.10 &               0.18 &           3769 &        144 &    10.0 &                             22 &           -7.95 &    0.5 &            0.0 &        5.0  & K \\
DR Tau                  &   10.8  &   0.46 &   0.61 &               0.86 &           4327 &        194 &    10.0 &                             21 &           -6.50 &    0.7 &            0.0 &        1.0  & V \\
DS Tau                  &   11.3  &   0.78 &   0.34 &               0.43 &           4040 &        158 &    11.2 &                             21 &           -7.89 &    0.6 &            0.0 &        2.0  & K \\
FN Tau                  &   11.8  &   1.27 &   1.40 &               0.14 &           4250 &        130 &     6.4 &                              7 &             - &    0.7 &            0.5 &       30.0  & K \\
FP Tau                  &   12.0  &   1.55 &   0.20 &               0.24 &           3486 &        128 &    27.9 &                             19 &           -9.45 &    0.3 &            0.0 &        1.0  & K \\
GH Tau                  &   11.5  &   1.41 &   0.40 &               0.79 &           3580 &        140 &    25.2 &                             19 &           -7.92 &    0.3 &            0.0 &        0.5  & K \\
GM Aur                  &   11.2  &   1.10 &   0.31 &               0.55 &           4338 &        158 &    12.6 &                             14 &           -8.02 &    0.9 &            0.2 &        5.0  & K \\
GO Tau                  &   13.1  &   1.10 &   1.20 &               0.08 &           3984 &        143 &    19.2 &                              4 &           -7.93 &    0.6 &            0.5 &       70.0  & K \\
HBC 374                 &   11.0  &   1.65 &   0.00 &               0.43 &           4007 &        125 &    12.9 &                             21 &             - &    0.6 &            0.0 &        2.0  & S \\
HBC 376                 &   11.4  &   1.08 &   0.00 &               0.28 &           4389 &        121 &    68.0 &                              5 &           -8.92 &    0.8 &            0.6 &       30.0  & S \\
HBC 407                 &   11.9  &   1.03 &   0.00 &               0.16 &           4649 &        125 &     8.8 &                              3 &             - &    0.7 &            0.6 &       70.0  & V \\
HBC 427                 &   10.7  &   1.21 &   0.20 &               0.87 &           4249 &        148 &    10.0 &                             21 &             - &    0.7 &            0.0 &        1.0  & K  \\
HD 285778               &    9.7  &   0.72 &   0.23 &               1.18 &           5358 &        119 &    17.6 &                              6 &           -8.03 &    1.3 &            1.0 &        7.0  & S \\
HP Tau                  &    -     &   1.55 &   0.39 &               0.29 &           3688 &        176 &   100.0 &                             20 &           -8.47 &    0.4 &            0.0 &        2.0  & V \\
IT Tau                  &   12.5  &   1.18 &   3.10 &               0.18 &           3913 &        161 &     - &                             13 &             - &    0.7 &            0.2 &       10.0  & K \\
LkCa 04                 &   11.3  &   1.38 &   0.35 &               0.40 &           3621 &        129 &    26.1 &                             19 &           -8.73 &    0.4 &            0.0 &        1.0  & K \\
LkCa 14                 &   10.8  &   1.21 &   0.00 &               0.58 &           4218 &        127 &    21.9 &                             22 &           -8.85 &    0.7 &            0.0 &        2.0 & K  \\
LkCa 15                 &   11.0  &   1.08 &   0.60 &               0.64 &           4201 &        158 &    12.5 &                             22 &           -8.87 &    0.7 &            0.0 &        2.0 & K  \\
LkCa 19                 &   10.0  &   0.96 &   0.00 &               1.63 &           4784 &        158 &    19.8 &                             16 &          -10.0 &    1.2 &            0.2 &        2.0  & K \\
RY Tau${}^{\rm (*)}$                  &    9.3  &   0.86 &   1.50 &              6.30 &           5750 &        133 &    48.8 &                             3 &           -7.11 &    1.8 &            1.7 &        5.0  & N \\
SU Aur                  &    8.9  &   0.58 &   0.90 &               4.31 &           4359 &        157 &    65.0 &                             20 &           -8.25 &    0.6 &            0.0 &        0.2  & V \\
UX Tau                  &   10.0  &   0.92 &   0.36 &               0.59 &           4427 &        139 &     9.9 &                             19 &           -9.00 &    0.9 &            0.1 &        3.0  & K \\
V1023 Tau               &   11.0  &   1.24 &   1.35 &               0.43 &           4007 &        125 &    12.9 &                             21 &           -7.78 &    0.6 &            0.0 &        2.0  & N \\
V1204 Tau               &   10.1  &   0.88 &   0.31 &               1.05 &           4800 &        138 &    22.5 &                              9 &             - &    1.2 &            0.6 &        5.0  & S \\
V1297 Tau               &   10.8  &   0.72 &   0.28 &               0.44 &           5039 &        117 &    17.5 &                              2 &             - &    0.9 &            0.9 &       50.0  & K \\
V1321 Tau               &   12.0  &   1.35 &   0.70 &               0.22 &           4042 &        146 &    11.2 &                              7 &             - &    0.7 &            0.5 &       30.0  & K \\
V1348 Tau               &   11.3  &   1.09 &   0.00 &               0.41 &           4441 &        155 &     4.8 &                             12 &             - &    0.9 &            0.3 &        7.0  & K \\
V1840 Ori               &   11.1  &   0.97 &   0.20 &               0.59 &           4617 &        149 &    14.5 &                             10 &             - &    1.0 &            0.5 &        7.0  & S \\
V830 Tau                &   11.1  &   1.22 &   0.30 &               0.44 &           4020 &        130 &    26.6 &                             21 &           -8.10 &    0.6 &            0.0 &        2.0  & K \\
V836 Tau                &   12.4  &   1.00 &   1.70 &               0.40 &           3631 &        168 &    13.4 &                             19 &           -8.98 &    0.4 &            0.0 &        1.0  & K \\
ZZ Tau                  &   11.9  &   1.22 &   1.00 &               0.16 &           4130 &        134 &    21.4 &                              7 &             - &    0.7 &            0.5 &       30.0  & K \\ \hline

\multicolumn{6}{l}{Orionis OB 1c association}   & & & & & & & & \\ \hline
HBC 167                 &   10.4  &   0.70 &   0.00 &               7.32 &           5504 &        406 &    18.0 &                             13 &             - &    2.0 &            0.6 &        1.0  & V \\ \hline

\multicolumn{6}{l}{Upper Scorpius association}  & & & & & & & & \\ \hline
1RXS J161951.4-215431   &   11.1  &   1.40 &   0.00 &                - &           3746 &       - &     - &                              2 &             - &    - &            - &        -  & K \\ \hline 

\multicolumn{6}{l}{Perseus molecular cloud}   & & & & & & & & \\ \hline
LkH$\rm \alpha$ 86          &   14.4  &   1.44 &   0.00 &               0.14 &           3657 &        321 &     6.8 &                             20 &             - &    0.4 &            0.0 &        5.0  & K \\
LRL 72&     - &   2.24 &   0.00 &               0.51 &           3488 &        249 &     9.3 &                             19 &             - &    0.3 &            0.0 &        0.7  & K \\ \hline

\multicolumn{6}{l}{AB Doradus moving group}  & & & & & & & & \\ \hline
HIP 17695 & 9.8 & 1.51 & 0.00 & 0.04 & 3349 & 16 & 18.0 & 19 & - & 0.2 & 0.0 & 20.0  & V \\ \hline

\multicolumn{6}{l}{$\rm \beta$ Pictoris moving group}   & & & & & & & & \\ \hline
HD 197481               &    7.4  &   1.46 &   0.04 &               0.10 &           3652 &          9 &     8.0 &                             21 &             - &    0.4 &            0.0 &       10.0  & V \\  \hline

\multicolumn{6}{l}{$\rm \eta$ Chamaeleontis cluster}   & & & & & & & & \\ \hline
RECX 04                 &   11.4  &   1.43 &   0.00 &               0.21 &           4023 &         99 &     6.0 &                             13 &             - &    0.7 &            0.2 &       10.0  & V \\
RECX 06                 &   12.4  &   1.42 &   0.00 &               0.10 &           3525 &         97 &    20.9 &                             19 &             - &    0.3 &            0.0 &        5.0  & V \\
RECX 07                 &   10.0  &   1.15 &   0.00 &               0.71 &           4325 &         98 &    30.0 &                             22 &             - &    0.8 &            0.0 &        3.0  & V \\
RECX 09                 &   12.8  &   1.46 &   0.00 &               0.10 &           3933 &         97 &     - &                              9 &          -10.4 &    0.6 &            0.3 &       30.0  & V \\
RECX 10                 &   11.3  &   1.42 &   0.00 &               0.21 &           4088 &         97 &     9.0 &                              7 &             - &    0.8 &            0.5 &       20.0  & V \\
RECX 11                 &   10.3  &   1.18 &   0.00 &               0.45 &           4392 &         98 &    13.0 &                             12 &           -9.77 &    0.9 &            0.3 &        7.0  & V \\
RECX 15                 &   12.7  &   0.83 &   0.00 &               0.06 &           3719 &         91 &    28.0 &                             11 &           -9.09 &    0.5 &            0.2 &       30.0  & V \\ \hline

\multicolumn{6}{l}{''Cha-Near'' region}   & & & & & & & & \\ \hline
RX J1147.7-7842         &   11.6  &   1.54 &   0.00 &               0.16 &           3881 &        106 &     - &                              9 &             - &    0.7 &            0.4 &       20.0  & V \\
RX J1204.6-7731         &   11.8  &   1.52 &   0.00 &               0.16 &           3584 &        100 &     6.0 &                             20 &             - &    0.4 &            0.0 &        3.0  & V \\ \hline

\multicolumn{6}{l}{TW Hydrae association}   & & & & & & & & \\ \hline
TWA 01                  &   10.0  &   0.89 &   0.27 &               0.26 &           4235 &         59 &    14.0 &                              9 &           -8.74 &    0.9 &            0.5 &       10.0  & V \\
TWA 06                  &   10.3  &   1.23 &   0.27 &               0.21 &           4268 &         65 &     - &                              7 &             - &    0.8 &            0.5 &       20.0  & V \\
TWA 07                  &   10.0  &   1.46 &   0.00 &               0.08 &           4017 &         33 &     4.4 &                              9 &             - &    0.6 &            0.3 &       30.0  & V \\
TWA 14                  &   11.3  &   1.27 &   0.10 &               0.20 &           3852 &         91 &     - &                             20 &             - &    0.6 &            0.0 &        7.0  & V \\
TWA 22                  &   11.3  &   1.73 &   0.00 &               0.01 &           2843 &         19 &     9.7 &                              2 &             - &    0.0 &            0.0 &        0.7  & V \\
TWA 23                  &   10.6  &   1.50 &   0.05 &               0.14 &           3508 &         55 &     - &                             19 &             - &    0.3 &            0.0 &        3.0  & K \\
TWA 25                  &    9.9  &   1.34 &   0.24 &               0.22 &           4019 &         53 &    12.9 &                              7 &             - &    0.8 &            0.5 &       20.0  & V \\  \hline \\
\end{longtable} 
\renewcommand{\tabcolsep}{6pt}
}

\begin{figure}[htbp]
\centering
 \includegraphics[clip, width=10cm, height=11cm]{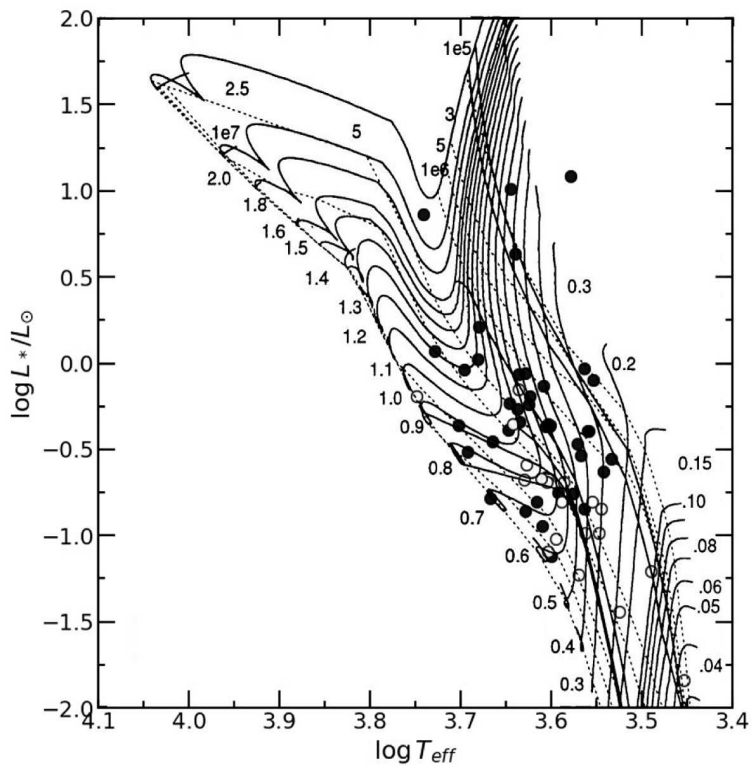} 
 \caption{HR diagram of the investigated PMS stars. The solid lines are Canuto \& Mazzitelli [CM] Alexander evolutionary tracks (\cite{dm94}). 
The filled circle symbols and the open circles indicate PMS stars in molecular clouds and moving groups, respectively.} 
 \label{fig:hrCMK}
\end{figure}

\figref{fig:hrCMK} presents the HR diagram of the investigated PMS stars.
The filled circle symbols indicate PMS stars in molecular clouds. The open circles represent PMS stars in the moving groups.
The luminosity, $T_{\rm eff}$, and distance of the objects were taken from {\it Gaia} DR2 (\cite{ba18}).
For objects, whose $T_{\rm eff}$ is not listed in {\it Gaia} DR2 (Bailer-Jones et al. 2018), we referred to other sources: \citet{palla} for AA Tau, \citet{kh95} for GH Tau, and \citet{pe13} for HD 197481, RECX 09, and TWA 22. 
The solid lines indicate Canuto \& Mazzitelli [CM] Alexander evolutionary tracks (\cite{dm94}). 
The stellar masses, $M_*$, masses of the bottom of the convective zone, $M_{\rm conv}$, and ages of the target stars were estimated from their evolutionary tracks. For example, objects with $M_{\rm conv} = 0 \, \mathrm{M_{\odot}}$ are fully convective. 

\section{Observations}

The observations were conducted with the Medium And Low Long slit Spectrograph (MALLS) mounted on the 2.0 m Nayuta Telescope at the Nishi-Harima Astronomical Observatory (NHAO), Japan. The data for four PMS stars was obtained between 2017 December 5 and 2019 February 6 with the 0.8'' width slit or the 1.2'' width slit using the $1800 \, \mathrm{l \cdot mm^{-1}}$ grating. These instrument settings achieved a wavelength coverage between $8350 \, \mathrm{\AA}$ and $9360 \, \mathrm{\AA}$ and a spectral resolution is between 7500 and 9000. The integration time for each object was between 600 s and 1200 s.

High-resolution spectroscopic observations of five PMS stars were conducted on September 18, 2007 with the High Dispersion Spectrograph (HDS; \cite{no02}) mounted on the Subaru Telescope. The data was obtained using the StdNIRb mode and with the 0.6'' width slit. These instrument settings achieved a wavelength coverage between $6650 \, \mathrm{\AA}$ and $9360 \, \mathrm{\AA}$ and a spectral resolution of 60,000. The integration time for each object was between 600 s and 1500 s. 

Archive data of 26 PMS stars and 5 standard stars obtained with HIRES mounted on the Keck Telescope was also used. The observer, date of the observations, wavelength coverage, and integration times are listed in \tabref{tab:arc}. The spectral resolution was 70,000. The archive data of six PMS stars obtained with the UVES ($R \sim 40,000$) mounted on the Very Large Telescope (VLT) was also used. The program IDs, principal investigators, and dates of the observations are listed in \tabref{tab:arc}. The wavelength coverage was between $5655 \, \mathrm{\AA}$ and $9496 \, \mathrm{\AA}$. The integration time for each object was between 10 s and 600 s. 
In addition, the archive data of 19 PMS stars obtained with the X-Shooter ($R \sim 8,000$) mounted on the VLT was also used. The wavelength coverage was between $5337 \, \mathrm{\AA}$ and $10200 \, \mathrm{\AA}$. The integration time for each object was between 2 s and 600 s. 

\begin{table}[h]
  \centering
  \caption{Details of the archive data from Keck/HIRES, VLT/UVES, and VLT/X-Shooter}
  \begin{tabular}{lllll} \hline
Program ID	  & 	PI-Observers			&	Observation Dates & Wavelength $\mathrm{[\AA]}$ & Integration Time $\mathrm{[s]}$\\ \hline
Keck / HIRES & {} & {} \\ \hline
C05H  &   N. Reid   &   1999-06-16 & 7600 - 9910 & 100 \\
C77H  &   L. Hillenbrand  &   1999-12-06, 07 & 6240 - 8680 & 1200 - 1500\\
C109H  &   L. Hillenbrand  &   2000-01-10, 11 & 6240 - 8680 & 300 - 900\\
C240Hr  &   S. E. Dahm  &   2007-11-30 & 4880 - 9380 & 460 - 1800\\
C095Hr  &   S. E. Dahm  &   2008-12-03, 04 & 4450 - 8900 & 10 - 1200\\
N107Hr  &   I. Pascucci  &   2012-12-01 & 4880 - 9380 & 300 - 3710\\
C252Hr  &   L. Hillenbrand  &   2013-12-26, 27 & 4880 - 9380 & 150 - 600\\
C247Hr  &   J. Carpenter  &   2015-06-01, 02 & 4880 - 9380 & 305 - 424\\ 
C250Hr  &   L. Hillenbrand  &   2016-05-17 & 4880 - 9380 & 22\\ 
C226Hr  &   L. Hillenbrand  &   2017-01-13 & 4880 - 9380 & 130\\ \hline

VLT / UVES & {} & {} & {} \\ \hline
075.C-0321(A)  &   M. Hempel  &   2005-08-26 & 5655 - 9496 & 200\\
082.C-0005(B)  &   A. Scholz   &   2008-10-02, 2009-01-08, 18 & 5655 - 9496 & 10 - 600 \\ \hline

VLT / X-Shooter & {} & {} & {} \\ \hline
084.C-1095(A)  &   G. Herczeg      &     2010-01-19, 2010-01-20 & 5337 - 10200 & 2 - 388\\
085.C-0238(A)   &  J. M. Alcala  &    2010-04-06, 07 & 5337 - 10200 & 20 - 480\\
086.C-0173(A)  &   J. M. Alcala   &   2010-01-12, 13 & 5337 - 10200 & 100 - 400\\
094.C-0327(A)  &   S. Alencar      &       2015-01-16 & 5337 - 10200 & 120\\
094.C-0805(A)  &   G. Herczeg       &    2015-01-15, 2015-03-07 & 5337 - 10200 & 24 - 600\\
094.C-0913(A)   &  C. F. Manara  &  2014-12-05 & 5337 - 10200 & 320 - 400\\ \hline
  \end{tabular}
　\label{tab:arc}
\end{table}

\section{Data Reduction}

The Image Reduction and Analysis Facility (IRAF) software package\footnote{IRAF is distributed by the National Optical Astronomy Observatories, which are operated by the Association of Universities for Research in Astronomy, Inc., under cooperative agreement with the National Science Foundation.} was used for data reduction. Overscan subtraction, dark subtraction, flat fielding, wavelength calibration using an Fe-Ne-Ar lamp, removal of scattered light, extraction of a spectrum, and continuum normalization were conducted for all the spectra obtained by MALLS.

The HDS data was reduced with overscan subtraction, bias subtraction, flat fielding, removal of scattered light, extraction of a spectrum, wavelength calibration using a Th-Ar lamp, and continuum normalization. A detailed description of the data reduction methods used herein is presented in Takagi et al. (2015). The HIRES data was reduced with the Mauna Kea Echelle Extraction (MAKEE) package. The UVES data and X-Shooter data had already been reduced.　

All object spectra were shifted to match the rest radial velocity. Therefore we measured the wavelengths of 2 un-blended Fe\,\emissiontype{I} absorption lines ($\lambda 8468.404, 8621.598 \, \mathrm{\AA}$), then shifted the wavelength of all spectra by the average of the difference between the measured wavelengths and laboratory wavelength.

The Ca\,\emissiontype{II} IRT emission lines are often located on the broad photospheric absorption lines of Ca\,\emissiontype{II}. In general, the photospheric absorption of the Ca\,\emissiontype{II} IRT line is strong, especially for K type stars.
And investigated PMS stars have late spectral type. An absorption line profile of a PMS spectrum could be obscured by continuum veiling. The amount of veiling, $V$, is defined as 
\begin{equation}
	V = \frac{W^0}{W} -1,
\end{equation}
where $W^0$ denotes the unveiled EQW and $W$ is the veiled EQW.

In this study, inactive stars with a spectral type similar to that of the target object were used as template stars (the inactive stars library, \cite{yee} ; \cite{pa18}). We obtained the Keck archive data of five inactive stars, named $\iota$ Psc (F7), 16 Cyg (G3), HD 166 (G8),  HD 88230 (K6), and GJ 412a (M1). For the correction of the rotational broadening, the spectra of the template stars were convolved with a Gaussian kernel to match the width of the absorption lines of each object. To estimate the amount of veiling, we measured the EQWs of Ti\,\emissiontype{I}, Fe\,\emissiontype{I}, and Cr\,\emissiontype{I} photospheric absorption lines between $8420$ and $8700 \, \mathrm{\AA}$ (\tabref{tab:abs_veiling}). 
In \tabref{tab:abs_veiling}, absorption lines of several stars could not be measured because of varying wavelength ranges of the individual data archives. 
If the amount of the veiling was estimated for more than five absorption lines, we calculated their mean values and residuals. \citet{basri90} claimed that the veiling is constant around $8500 \, \mathrm{\AA}$. We assumed that each of the Ca\,\emissiontype{II} IRT emission line has the same veiling value as all the others. We substituted the EQW of the lines of the PMS spectrum for $W$, that of the template star for $W^0$, and then obtained the amount of veiling for each absorption line. The average value of $V$ and standard deviation $\sigma_V$ were calculated (\tabref{tab:result_of_obs}). In case $V - \sigma$ is negative or the veiling values were measured for less than five absorption lines, we regarded $V$ as $0$. 

\begin{table}[h]
  \centering
  \caption{Absorption lines for estimating the amount of veiling.}
  \begin{tabular}{ll} \hline
{} & Wavelength $\, \mathrm{[\AA]}$ \\ \hline
Ti\,\emissiontype{I} & 8426.497\\
Ti\,\emissiontype{I} & 8450.871\\
Cr\,\emissiontype{I} & 8455.288\\
Fe\,\emissiontype{I} & 8468.404\\
Fe\,\emissiontype{I} & 8621.598\\
Fe\,\emissiontype{I} & 8632.412\\
Ti\,\emissiontype{I} & 8682.988\\
Fe\,\emissiontype{I} & 8688.600\\
Ti\,\emissiontype{I} & 8692.326\\ \hline
  \end{tabular}
　\label{tab:abs_veiling}
\end{table}

For objects indicating a significant veiling value, we made a veiled spectrum of the template star.
First, we added the veiling value $V$ to the normalized spectrum of the template star.
Then, the continuum component was normalized again to unity by dividing by $(1+V)$.
The veiled spectrum of the template star was subtracted from that of the target star indicating a significant veiling value.
For objects indicating $V = 0$, the spectrum of the template star which was not veiled was subtracted.

\begin{figure}[h]
\begin{center}
 \includegraphics[width=10cm]{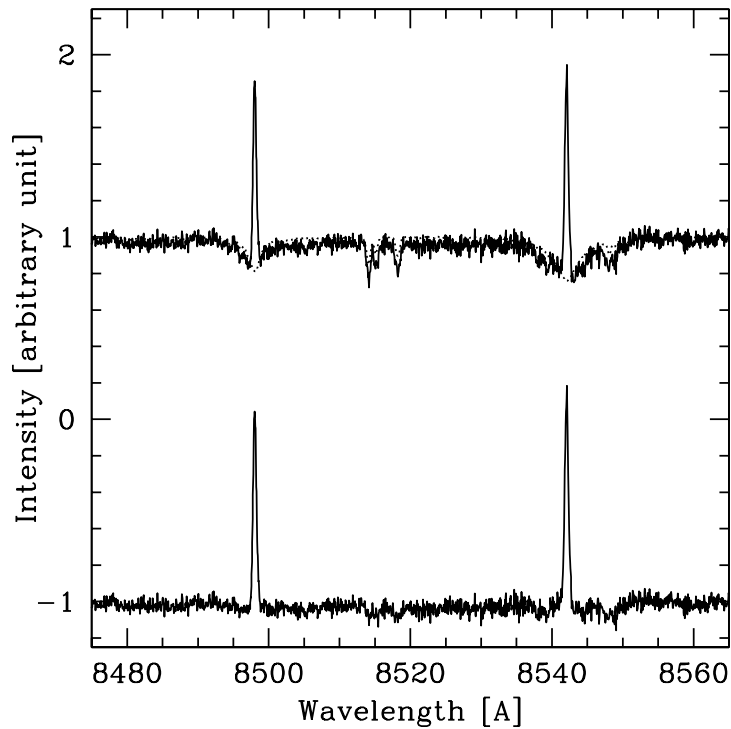}
 \vspace{0.3cm}
 \caption{The emission profiles of the Ca\,\emissiontype{II} IRT lines ($\lambda 8498, 8542 \, \mathrm{\AA}$). The observed spectrum of AA Tau is shown in the top of the panel with a solid line.
The dotted line is the fitted inactive star, a K6 template star.
The difference between the observed and template spectra is shown in the bottom of the panel, where the Ca\,\emissiontype{II} IRT narrow lines appear in the emission.
The difference between the AA Tau spectrum and the template star's spectrum is shown shifted by -1.0 for display purposes. Note that all of the nine absorption lines used for estimating the amount of the veiling (\tabref{tab:abs_veiling}) are out of range for this figure.}
 \label{fig:reduction_abs}
\end{center}
\end{figure}

\figref{fig:reduction_abs} shows the procedures of the spectral subtraction of the Ca\,\emissiontype{II} IRT lines of  AA Tau. 
The solid line at the top of the panel shows the observed spectrum of AA Tau before subtraction of the spectrum of the template star.
In this case, the Ca\,\emissiontype{II} IRT lines display emission profiles over a broad absorption feature.
The dotted line is the fitted spectrum of the veiled template star. 
Continuum components of both spectra were normalized to unity.
The solid line in the bottom of the panel represents the observed spectrum of AA Tau after subtracting the spectrum of the template star.
Any broad absorption feature was removed completely, and only the emission component remained.
Unfortunately, any absorption lines (\tabref{tab:abs_veiling}) to measure the veiling value are not shown in \figref{fig:reduction_abs}.
The spectra of most objects only contain emission components after this subtraction has been performed.
The subtraction of the template spectrum is necessary for a correct measurement of the EQWs for the Ca\,\emissiontype{II} IRT emission lines.

Before measuring the EQWs, the continuum component of the spectra was added to unity. To obtain the EQWs of the Ca\,\emissiontype{II} IRT emission lines, the area of the emission profile was directly integrated. We also measured their full width at half maximums (FWHMs) by fitting with a Gaussian function. The EQW errors were estimated by multiplying the standard deviation of the continuum by the wavelength range of the emission line. The wavelength range for measuring standard deviation is $\lambda 8483-8492 \, \mathrm{\AA}$ for the Ca\,\emissiontype{II} $\lambda 8498, 8542 \, \mathrm{\AA}$ lines and $\lambda 8623-8632 \, \mathrm{\AA}$ for the Ca\,\emissiontype{II} $\lambda 8662$ line. These ranges are free from any emission and absorption lines.
Further for our discussion we only consider emission profiles the S/N larger than 3, for calculating EQWs, FWHMs, etc. 

\clearpage
\chapter{Results}
\label{result}

\begin{figure}[htbp]
\begin{center}
    \vspace{-2.5cm}
    \includegraphics[clip, width=16cm]{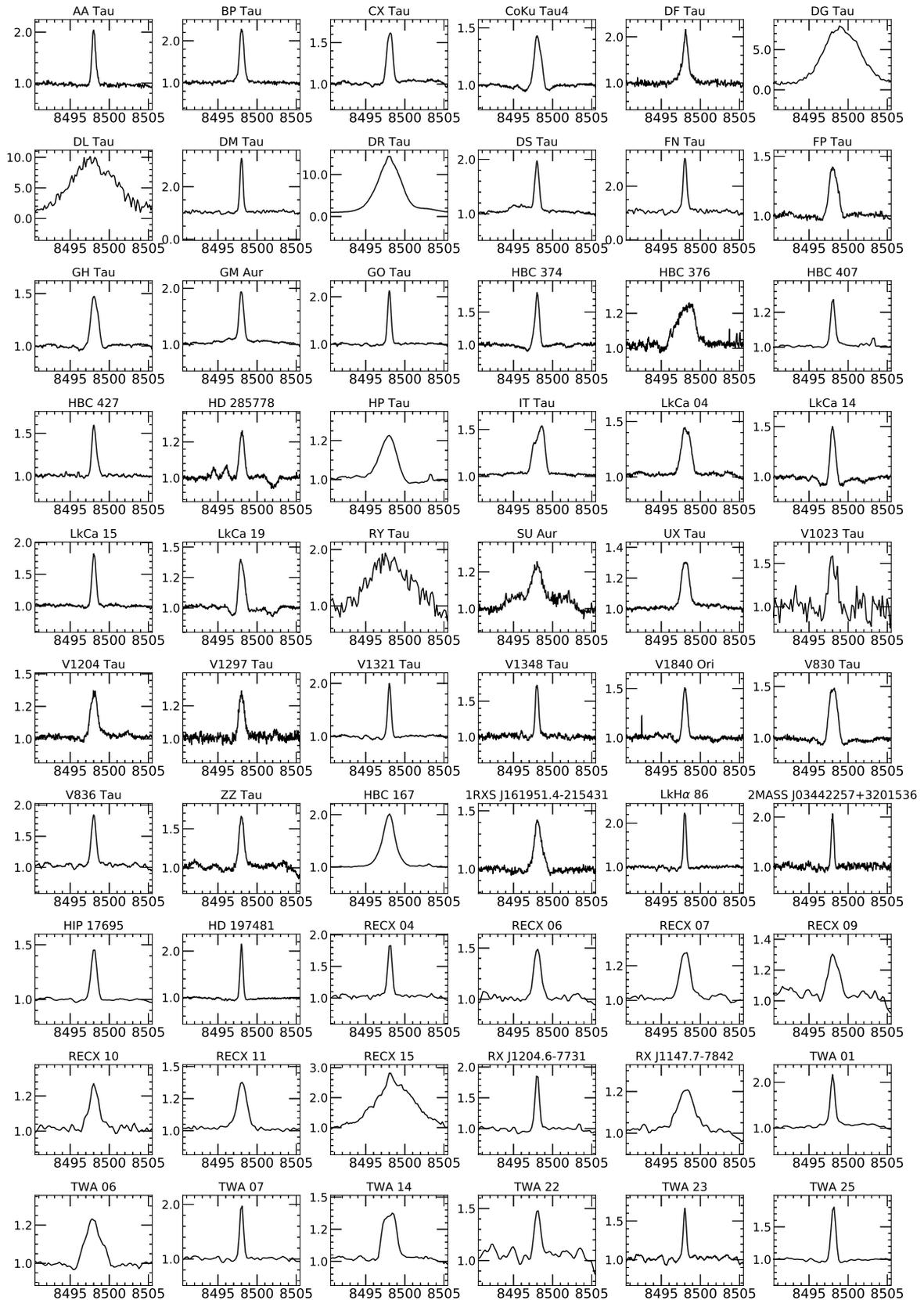}
    \vspace{-29cm}
    \caption{Ca\,\emissiontype{II} IRT emission lines ($\lambda 8498 \, \mathrm{\AA}$) of all the PMS stars. 
The spectra are normalized to unity.
Photospheric absorption lines have already been subtracted.
}
    \label{fig:result_emi}
\end{center}
\end{figure}

The EQWs of the Ca\,\emissiontype{II} IRT emission lines and their errors are listed in \tabref{tab:result_of_obs}.  Seven objects have broad emission lines of Ca\,\emissiontype{II} $\lambda 8498 \, \mathrm{\AA}$ (FWHM $> 100 \, \mathrm{km \cdot s^{-1}}$), while most PMS stars exhibit narrow emission lines (FWHM $\leq 100 \, \mathrm{km \cdot s^{-1}}$). \figref{fig:result_emi} shows the spectra of the $\lambda 8498 \, \mathrm{\AA}$ emission line after subtracting the photospheric absorption. The emission lines of DG Tau, DL Tau, and DR Tau are broad and strong ($W_{\rm IRT} \sim 50 \, \mathrm{\AA}$), while those of RY Tau, SU Aur, RECX 15, and RX J1147.7-7842 are broad but not strong ($W_{\rm IRT} < 10 \, \mathrm{\AA}$). All EQWs of narrow emission lines are weaker than $5 \, \mathrm{\AA}$. We note that a part of the narrow emission component of the PMS stars could be buried by the photospheric absorption before removal. In our sample, most objects that belong to the moving group only have absorption lines before removal of the photospheric absorption.

We obtained the amount of the veiling for 27 objects. Among them the amount of veiling was measured in the previous studies for 13 objects. The correlation coefficient between the literature values and our measured values is 0.84.  The literature value was found to be within the uncertainty of the measured $V$ for 8 objects. We note that on multiple occasions the variation of the veiling value have often been reported in several literature (\cite{basri90}; \cite{bb96}; Hartigan et al. 1989, 1991). 


\begin{small}

\begin{longtable}{p{30mm}p{20mm}p{20mm}p{20mm}p{7mm}p{7mm}p{7mm}p{20mm}}
\caption{EQWs, FWHMs of the Ca\emissiontype{II} IRT emission lines ($\lambda$ 8498, 8542, 8662 $A$) and the veiling value.}
\label{tab:result_of_obs}
\hline
Object & \multicolumn{3}{c}{ $W_{\rm IRT} \, \mathrm{[\AA]}$ }  &  \multicolumn{3}{c}{ ${\rm FWHM} \, \mathrm{[km \cdot s^{-1}}$] }  &  Veiling \\

{}   & $\lambda 8498$  &  $\lambda 8542$   &  $\lambda 8662$     & $\lambda 8498$  &  $\lambda 8542$   &  $\lambda 8662$    &         \\
\hline
\endhead
\endfoot

\endlastfoot
\multicolumn{2}{l}{Taurus-Auriga molecular cloud}        & &     &     & &         &      \\ \hline
AA Tau                  &      0.65 $\pm$ 0.04 &      0.77 $\pm$ 0.04 &      0.61 $\pm$ 0.04 &                                              20 &                                              23 &                                              20 &    1.29 $\pm$ 1.20 \\
BP Tau                  &      1.05 $\pm$ 0.07 &      1.68 $\pm$ 0.12 &      1.15 $\pm$ 0.11 &                                              24 &                                              32 &                                              26 &    0.56 $\pm$ 0.40 \\
CX Tau                  &      0.59 $\pm$ 0.03 &      0.77 $\pm$ 0.04 &      0.72 $\pm$ 0.05 &                                              31 &                                              35 &                                              36 &   -0.11 $\pm$ 0.20 \\
CoKu Tau4               &      0.53 $\pm$ 0.03 &      0.83 $\pm$ 0.04 &       0.60 $\pm$ 0.04 &                                              37 &                                              42 &                                              38 &     - \\
DF Tau                  &      1.14 $\pm$ 0.15 &      2.02 $\pm$ 0.25 &      1.38 $\pm$ 0.19 &                                              33 &                                              55 &                                              45 &     - \\
DG Tau                  &     49.49 $\pm$ 1.58 &     46.83 $\pm$ 1.74 &      51.39 $\pm$ 2.50 &                                             185 &                                             221 &                                             211 &     - \\
DL Tau                  &     53.88 $\pm$ 8.14 &     50.65 $\pm$ 9.21 &     39.37 $\pm$ 7.89 &                                             249 &                                             271 &                                             250 &     - \\
DM Tau                  &      1.02 $\pm$ 0.06 &      1.29 $\pm$ 0.09 &      1.07 $\pm$ 0.09 &                                              16 &                                              22 &                                              19 &  -0.19 $\pm$ 0.18 \\
DR Tau                  &      53.1 $\pm$ 0.73 &     59.83 $\pm$ 0.75 &      49.03 $\pm$ 0.30 &                                             126 &                                             163 &                                             153 &   2.64 $\pm$ 1.48 \\
DS Tau                  &      0.75 $\pm$ 0.04 &      1.19 $\pm$ 0.06 &      1.05 $\pm$ 0.04 &                                              26 &                                              35 &                                              35 &   0.35 $\pm$ 0.32 \\
FN Tau                  &       1.32 $\pm$ 0.10 &        - &      1.49 $\pm$ 0.14 &                                              20 &                                               - &                                              28 &  -0.18 $\pm$ 0.29 \\
FP Tau                  &      0.55 $\pm$ 0.04 &        - &      0.55 $\pm$ 0.06 &                                              45 &                                               - &                                              41 &     - \\
GH Tau                  &      0.62 $\pm$ 0.04 &      0.72 $\pm$ 0.04 &      0.56 $\pm$ 0.03 &                                              38 &                                              40 &                                              37 &     - \\
GM Aur                  &      0.73 $\pm$ 0.02 &      1.41 $\pm$ 0.02 &      1.15 $\pm$ 0.02 &                                              27 &                                              42 &                                              38 &   0.31 $\pm$ 0.16 \\
GO Tau                  &      0.62 $\pm$ 0.04 &      0.86 $\pm$ 0.05 &      0.72 $\pm$ 0.05 &                                              19 &                                              23 &                                              22 &   -0.20 $\pm$ 0.16 \\
HBC 374                 &      0.57 $\pm$ 0.03 &      0.97 $\pm$ 0.06 &        - &                                              23 &                                              33 &                                               - &     - \\
HBC 376                 &      0.67 $\pm$ 0.08 &       0.90 $\pm$ 0.08 &        - &                                              95 &                                             111 &                                               - &     - \\
HBC 407                 &      0.26 $\pm$ 0.01 &      0.33 $\pm$ 0.01 &      0.27 $\pm$ 0.01 &                                              24 &                                              27 &                                              25 &     - \\
HBC 427                 &      0.44 $\pm$ 0.02 &      0.58 $\pm$ 0.03 &      0.52 $\pm$ 0.04 &                                              25 &                                              32 &                                              31 &   0.07 $\pm$ 0.17 \\
HD 285778               &      0.24 $\pm$ 0.02 &      0.39 $\pm$ 0.04 &        - &                                              26 &                                              41 &                                               - &     - \\
HP Tau                  &      0.56 $\pm$ 0.04 &      0.84 $\pm$ 0.05 &      0.72 $\pm$ 0.03 &                                              84 &                                             110 &                                              90 &     - \\
IT Tau                  &      0.81 $\pm$ 0.03 &      1.05 $\pm$ 0.03 &      0.88 $\pm$ 0.06 &                                              55 &                                              61 &                                              58 &   0.12 $\pm$ 0.23 \\
LkCa 04                 &      0.62 $\pm$ 0.05 &      0.92 $\pm$ 0.06 &      0.83 $\pm$ 0.07 &                                              51 &                                              68 &                                              67 &     - \\
LkCa 14                 &      0.54 $\pm$ 0.04 &      0.74 $\pm$ 0.04 &      0.62 $\pm$ 0.04 &                                              29 &                                              32 &                                              29 &     - \\
LkCa 15                 &      0.56 $\pm$ 0.02 &      0.58 $\pm$ 0.02 &      0.56 $\pm$ 0.05 &                                              22 &                                              26 &                                              23 &   0.25 $\pm$ 0.17 \\
LkCa 19                 &      0.37 $\pm$ 0.02 &      0.51 $\pm$ 0.03 &      0.38 $\pm$ 0.03 &                                              32 &                                              36 &                                              31 &    0.40 $\pm$ 0.18 \\
RY Tau                  &      6.53 $\pm$ 1.57 &      4.94 $\pm$ 1.47 &      4.21 $\pm$ 1.17 &                                             222 &                                             220 &                                             170 &     - \\
SU Aur                  &      1.11 $\pm$ 0.16 &      1.87 $\pm$ 0.19 &      1.51 $\pm$ 0.26 &                                             117 &                                             179 &                                             113 &     - \\
UX Tau                  &      0.37 $\pm$ 0.02 &      0.49 $\pm$ 0.02 &      0.43 $\pm$ 0.03 &                                              40 &                                              41 &                                              45 &     - \\
V1023 Tau               &      0.88 $\pm$ 0.24 &      1.32 $\pm$ 0.29 &        - &                                              43 &                                              53 &                                               - &     - \\
V1204 Tau               &      0.37 $\pm$ 0.04 &      0.39 $\pm$ 0.03 &        - &                                              35 &                                              36 &                                               - &     - \\
V1297 Tau               &       0.30 $\pm$ 0.04 &      0.36 $\pm$ 0.04 &      0.28 $\pm$ 0.05 &                                              29 &                                              30 &                                              28 &     - \\
V1321 Tau               &      0.54 $\pm$ 0.02 &      0.73 $\pm$ 0.04 &      0.62 $\pm$ 0.04 &                                              19 &                                              24 &                                              22 &  -0.32 $\pm$ 0.16 \\
V1348 Tau               &       0.40 $\pm$ 0.04 &      0.54 $\pm$ 0.05 &      0.42 $\pm$ 0.05 &                                              18 &                                              25 &                                              20 &   0.31 $\pm$ 0.33 \\
V830 Tau                &      0.75 $\pm$ 0.04 &      1.17 $\pm$ 0.06 &      0.79 $\pm$ 0.04 &                                              43 &                                              46 &                                              40 &  -0.37 $\pm$ 0.14 \\
V836 Tau                &      0.68 $\pm$ 0.05 &      0.84 $\pm$ 0.05 &      0.81 $\pm$ 0.06 &                                              28 &                                              40 &                                              34 &   0.08 $\pm$ 0.35 \\
ZZ Tau                  &      0.62 $\pm$ 0.07 &        0.84 $\pm$ 0.07 &      0.56 $\pm$ 0.08 &                                              28 &                                               33 &                                              35 &     - \\ \hline

\multicolumn{2}{l}{Orionis OB 1c association}       & &     &     & &         &      \\ \hline
HBC 167                 &      1.93 $\pm$ 0.05 &      3.28 $\pm$ 0.08 &      2.43 $\pm$ 0.05 &                                              67 &                                              85 &                                              80 &   0.13 $\pm$ 0.26 \\ \hline

\multicolumn{2}{l}{Upper Scorpius association}        & &     &     & &         &      \\ \hline
1RXS J161951.4-215431   &      0.55 $\pm$ 0.06 &      0.84 $\pm$ 0.07 &      0.69 $\pm$ 0.06 &                                              43 &                                              52 &                                              43 &     - \\ \hline

\multicolumn{2}{l}{Perseus molecular cloud}        & &     &     & &         &      \\ \hline
LkH$\rm \alpha$ 86          &      0.62 $\pm$ 0.04 &      0.85 $\pm$ 0.06 &      0.68 $\pm$ 0.06 &                                              15 &                                              20 &                                              18 &  -0.24 $\pm$ 0.18 \\
LRL 72 &      0.47 $\pm$ 0.04 &      0.58 $\pm$ 0.08 &      0.44 $\pm$ 0.07 &                                              15 &                                              20 &                                              21 &  -0.21 $\pm$ 0.19 \\ \hline

\multicolumn{2}{l}{AB Doradus moving group}        & &     &     & &         &      \\ \hline
HIP 17695               &      0.42 $\pm$ 0.03 &      0.58 $\pm$ 0.06 &      0.41 $\pm$ 0.07 &                                              29 &                                              33 &                                              32 &     - \\ \hline

\multicolumn{2}{l}{$\rm \beta$ Pictoris moving group}        & &     &     & &         &      \\ \hline
HD 197481               &      0.57 $\pm$ 0.02 &      0.69 $\pm$ 0.03 &      0.56 $\pm$ 0.03 &                                              15 &                                              19 &                                              18 &     - \\ \hline 

\multicolumn{2}{l}{$\rm \eta$ Chamaeleontis cluster}        & &     &     & &         &      \\ \hline
RECX 04                 &      0.61 $\pm$ 0.07 &      0.84 $\pm$ 0.11 &      0.63 $\pm$ 0.08 &                                              23 &                                              31 &                                              25 &   0.08 $\pm$ 0.38 \\
RECX 06                 &      0.59 $\pm$ 0.08 &      0.71 $\pm$ 0.12 &       0.51 $\pm$ 0.10 &                                              38 &                                              43 &                                              40 &     - \\
RECX 07                 &      0.58 $\pm$ 0.05 &      0.83 $\pm$ 0.07 &      0.67 $\pm$ 0.04 &                                              54 &                                              58 &                                              59 &     - \\
RECX 09                 &      0.49 $\pm$ 0.11 &      0.92 $\pm$ 0.15 &      0.66 $\pm$ 0.11 &                                              57 &                                              85 &                                              78 &     - \\
RECX 10                 &      0.52 $\pm$ 0.08 &      0.67 $\pm$ 0.09 &      0.51 $\pm$ 0.07 &                                              54 &                                              62 &                                              52 &   0.02 $\pm$ 0.21 \\
RECX 11                 &      0.59 $\pm$ 0.03 &      0.87 $\pm$ 0.05 &      0.72 $\pm$ 0.05 &                                              52 &                                              63 &                                              61 &   0.41 $\pm$ 0.83 \\
RECX 15                 &      9.28 $\pm$ 0.58 &      9.46 $\pm$ 0.58 &      7.32 $\pm$ 0.52 &                                             204 &                                             226 &                                             222 &    0.11 $\pm$ 0.40 \\ \hline \\

\multicolumn{2}{l}{''Cha-Near'' region}        & &     &     & &         &      \\ \hline
RX J1147.7-7842         &      0.57 $\pm$ 0.08 &      0.94 $\pm$ 0.09 &      0.63 $\pm$ 0.05 &                                             104 &                                             112 &                                              94 &     - \\
RX J1204.6-7731         &      0.64 $\pm$ 0.05 &      0.87 $\pm$ 0.07 &      0.64 $\pm$ 0.05 &                                              21 &                                              27 &                                              22 &     - \\ \hline

\multicolumn{2}{l}{TW Hydrae association}        & &     &     & &         &      \\ \hline
TWA 01                  &      1.03 $\pm$ 0.07 &      1.96 $\pm$ 0.08 &       1.72 $\pm$ 0.10 &                                              26 &                                              38 &                                              35 &   0.37 $\pm$ 0.55 \\
TWA 06                  &      0.56 $\pm$ 0.03 &       0.90 $\pm$ 0.04 &      0.72 $\pm$ 0.05 &                                              80 &                                              85 &                                              83 &     - \\
TWA 07                  &      0.62 $\pm$ 0.04 &      0.73 $\pm$ 0.05 &      0.55 $\pm$ 0.06 &                                              19 &                                              24 &                                              20 &     - \\
TWA 14                  &      0.75 $\pm$ 0.05 &      1.15 $\pm$ 0.07 &      0.85 $\pm$ 0.07 &                                              66 &                                              81 &                                              76 &   0.03 $\pm$ 0.33 \\
TWA 22                  &      0.55 $\pm$ 0.12 &      1.06 $\pm$ 0.18 &       0.80 $\pm$ 0.18 &                                              32 &                                              56 &                                              74 &     - \\
TWA 23                  &      0.36 $\pm$ 0.04 &      0.48 $\pm$ 0.06 &      0.29 $\pm$ 0.04 &                                              18 &                                              22 &                                              19 &  -0.22 $\pm$ 0.19 \\
TWA 25                  &      0.62 $\pm$ 0.02 &      0.82 $\pm$ 0.03 &      0.72 $\pm$ 0.03 &                                              24 &                                              27 &                                              26 &     - \\ \hline
\end{longtable}

\end{small}

\clearpage
\chapter{Discussion}
\label{discussion}

\section{Emission Line Ratios of the Ca\,\emissiontype{II} IRT Lines} 
\label{thiswork : tauc}

The ratios of the EQWs, $W_{\lambda 8542} / W_{\lambda 8498}$ and $W_{\lambda 8662} / W_{\lambda 8498}$ are sensitive to the conditions of the emitting plasma. It was shown that solar chromospheric plages and flares have the ratio of $1 \leq W_{\lambda 8542} / W_{\lambda 8498} \leq 2$ (\cite{hs80}). 
In contrast, optically thin emission sources, such as solar prominences have a high value of this ratio, $W_{\lambda 8542} / W_{\lambda 8498} > 4$.
In several studies (\cite{hs80} ; \cite{hp92}), TTSs showed small line ratios of the Ca\,\emissiontype{II} IRT emission lines, indicating dense chromospheric region such as plages and flares. 

\begin{figure}[htbp]
\begin{center}
  \includegraphics[keepaspectratio, width=10.5cm]{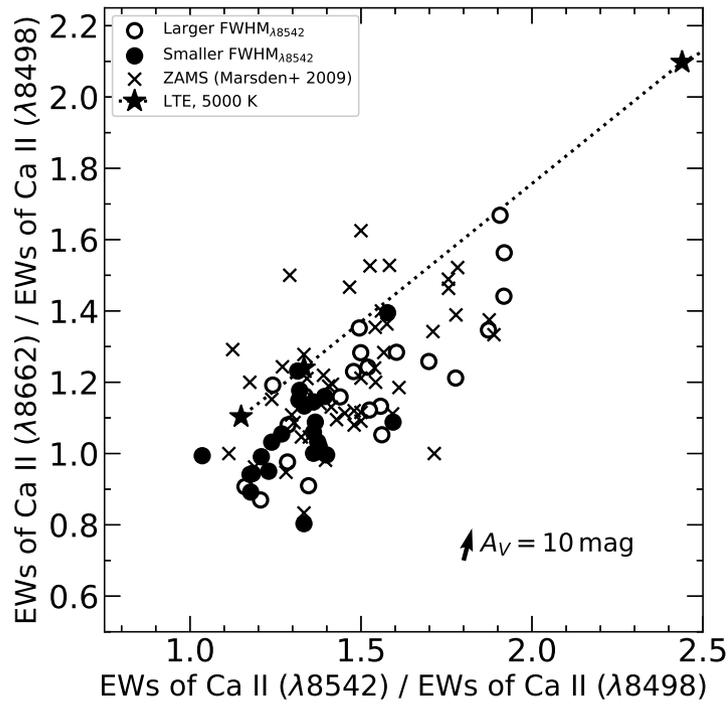} 
  \caption{Ratios of the EQWs of the Ca\,\emissiontype{II} IRT emission lines.
The circle symbols represent the PMS stars in the molecular clouds and moving groups.
The PMS stars are separated into two groups by the median of their ${\rm FWHM}_{\lambda 8542}$; the open circles show the group of the PMS stars with a broader ${\rm FWHM}_{\lambda 8542}$ than the median. 
The filled circles represent the narrower group. 
The low-mass stars in the young open clusters IC 2391 and IC 2602 are plotted as cross symbols (\cite{marsden09}).
The dotted line represents isothermal slabs ($T = 5000 \, \mathrm{K}$) in LTE (\cite{hs80} ; \cite{hp92}).
The black star symbols on the line correspond to slab optical depths of 10, 100, and 1000, beginning from the upper right.
The arrow in the lower right is the redding vector with $A_V = 10 \, \mathrm{mag}$.}
  \label{fig:EQWratio2}
\end{center}
\end{figure}

The ratios of the EQWs of the Ca\,\emissiontype{II} IRT emission lines are plotted in \figref{fig:EQWratio2}. 
The circle symbols represent the PMS stars in the molecular clouds and the moving groups. The PMS stars are separated into two groups by the median of their ${\rm FWHM}_{\lambda 8542}$; the open circles represent the group of the PMS stars with a broader ${\rm FWHM}_{\lambda 8542}$ than the median. The filled circles represent the narrower group. For comparison, we also plotted the low-mass stars in the young open clusters IC 2391 and IC 2602 studied in \citet{marsden09} using cross symbols. The dotted line indicates isothermal slabs ($T = 5000 \, \mathrm{K}$) in LTE (\cite{hs80} ; \cite{hp92}). The black star symbols on the line correspond to the slab optical depth of 10, 100, and 1000. The arrow in the lower right is the redding vector calculated by the equation below (\cite{allen}) with $A_V = 10 \, \mathrm{mag}$,
\begin{eqnarray}
\frac{A_{\lambda}}{A_V} & = & 0.41 \times (\lambda [\mu m])^{-1.75}.
\end{eqnarray}

As shown in \figref{fig:EQWratio2}, the values of $W_{\lambda 8542} / W_{\lambda 8498}$ are 1.0 - 2.0 and $W_{\lambda 8662}/W_{\lambda 8498}$ are 0.7 - 1.7 for both, the PMS samples and cluster members. This suggests that the emission originates from the regions such as chromospheric plages and flares. Moreover, there is a tendency that emission lines showing relatively large $W_{\lambda 8542}/W_{\lambda 8498}$ are broad. The large FWHM of the line suggests a large turbulent velocity (\cite{hp92}).

Our results for the PMS stars are consistent with \citet{hs80}, \citet{hp92}, and \citet{f17}. 
\citet{f17} investigated the ratio of the line flux, $F_{\lambda 8542}/F_{\lambda 8498}$, of Class II and Class III objects in the Lupus star forming region.
These objects have a low ratio of $1 < F_{\lambda 8542}/F_{\lambda 8498} < 2$.
Not only the PMS stars in our sample but also the cluster members in IC 2391 and IC 2602 show a ratio of $W_{\lambda 8542}/W_{\lambda 8498}$ of 1.0 - 2.0, indicating that the Ca\,\emissiontype{II} IRT emission lines are emitted from the regions analogous to solar plages and flares.

The relationship between optical thickness and line broadening has been studied for the Ca\,\emissiontype{II} HK emission lines. Hamann \& Persson (1989, 1992a) noticed that the Ca\,\emissiontype{II} HK emission lines of the TTSs are stronger and broader than the Ca\,\emissiontype{II} IRT emission lines. They interpreted that the Ca\,\emissiontype{II} HK emission lines are optically thicker than the Ca\,\emissiontype{II} IRT emission lines because the Ca\,\emissiontype{II} HK emission lines are generally stronger than the Ca\,\emissiontype{II} IRT emission lines. 
\citet{cg87} calculated the Ca\,\emissiontype{II} K line profile in the non-LTE chromospheric model. According to this model, with a small chromospheric mass column density, the Ca\,\emissiontype{II} K line shows a narrow and weak emission. In case the mass column density increases, the Ca\,\emissiontype{II} K line exhibits not only stronger but also broader emissions. 
\citet{bb93} constructed photospheric and chromospheric models to match the Ca\,\emissiontype{II} HK and IRT emission line profiles of six PMS stars. They found that the lines are shaped by mass and temperature of the photosphere, the chromospheric temperature and microtublence, and the temperature gradient of the lower chromosphere. A broad absorption line with a narrow emission core at the line center was successfully reproduced in both CTTSs and WTTSs with the chromospheric models.

\section{Chromospheric Activity and Mass Accretion Rate}
\label{thiswork : mass_acc_weak}

We compared the strengths of the Ca\,\emissiontype{II} IRT emission lines and mass accretion rates to discuss whether the chromosphere is activated by mass accretion from the protoplanetary disk.

\citet{mo05} investigated the chromospheric activity of CTTSs, very low-mass young stars ($0.075 \leq M_* <  0.15 \, \mathrm{M_{\odot}}$) and young brown dwarfs ($M_* \leq 0.075 \, \mathrm{M_{\odot}}$). They selected ''accretors'' by applying a number of accretion diagnostics, such as an H${\rm \alpha} 10 \%$ width $\geq 200 \, \mathrm{km \cdot s^{-1}}$. For the accretors, the surface flux of the Ca\,\emissiontype{II} emission line, $F^{\prime}_{\lambda 8662}$, showed a positive correlation with their mass accretion rate, $\dot{M}$, for approximately 4 orders of magnitude. Hence, they claim that the mass accretion rate can be estimated using the strength of the broad components of the Ca\,\emissiontype{II} IRT emission lines. In \citet{mo05}, the mass accretion rates of CTTSs were taken from \citet{mu98} and \citet{wb03}, in which the rates were estimated from the amount of the veiling in the $U$-, $V$- and $R_C$-bands. For the very low-mass young stars and the young brown dwarfs, the mass accretion rates were taken from \cite{mu03}, in which the rates were estimated from EQWs of the H$\mathrm{\alpha}$ emission lines. 

For calculating $F^{\prime}_{\rm IRT}$, a bolometric continuum flux per unit area at a stellar surface, $F$, was calculated at first.
We used the $i$-band mag (the AB system) of the UCAC4 Catalogue (\cite{za13}), the stellar radius, and the distance of the objects (\tabref{tab:objects1}). 
$F$ is given as
\begin{eqnarray}
	\log \frac{f}{f_0} & = & - \frac{2}{5} \times (m_{i*} - A_I),\\
	F & = & f \times \left( \frac{d}{R_*} \right)^2,
\end{eqnarray}
where $f$ is the bolometric continuum flux of the object per unit area as observed on Earth. 
$m_{i*}$ is the apparent magnitude of the object in the $i$-band.
The bolometric continuum flux per unit area under $m_i = 0 \, \mathrm{mag}$ (the AB system) condition, $f_0$, is $1.852 \times 10^{-12} \, \mathrm{W \cdot m^{-2} \cdot A^{-1}}$ (\cite{f96}).
$A_I$ is the absorption coefficient for $I$ mag, which is established with the absorption coefficient for $V$ mag, $A_V$ (\cite{v68}): 
\begin{equation}
	A_I = A_V \times 0.482.
\end{equation}
$d$ denotes the distance from an object to Earth ({\it Gaia} DR2: \cite{ba18}). 
Unfortunately, the distance from GH Tau is not listed in {\it Gaia} DR2, so we substituted that of the Taurus molecular cloud ($140 \, \mathrm{pc}$).
$R_*$ is the stellar radius estimated using Stefan-Boltzmann's law with the photospheric luminosity, $T_{\rm eff}$, and the distance of the objects in {\it Gaia} DR2 (\cite{g18}, \cite{ba18}).
For object whose $T_{\rm eff}$ is not listed in {\it Gaia} DR2, we used the luminosity and $T_{\rm eff}$ listed in other papers: \citet{palla} for AA Tau, \citet{kh95} for GH Tau, \citet{pe13} for HD 197481, RECX 09, and TWA 22.
$F$ was multiplied by the EQW of the Ca\,\emissiontype{II} IRT lines.
\begin{equation}
	F^{\prime}_{\rm IRT} = F \times W_{\rm IRT}, 
\end{equation}

\begin{figure}[h]
\begin{center}
  \includegraphics[width=11cm]{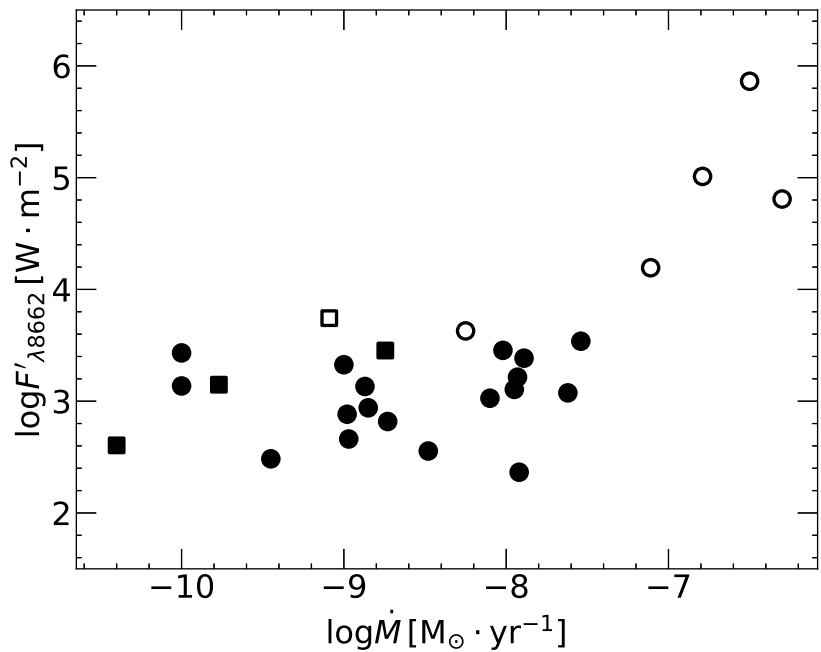}
  \caption{Surface flux of the Ca\,\emissiontype{II} emission line, $F^{\prime}_{\lambda 8662}$, as a function of mass accretion rate, $\dot{M}$. The circle symbols represent the PMS stars in the molecular clouds and the square symbols are the PMS stars in the moving groups. The open symbols show the objects with a broad Ca\,\emissiontype{II} IRT emission line (FWHM $> 100 \, \mathrm{km \cdot s^{-1}}$) and the filled symbols represent the objects with a narrow Ca\,\emissiontype{II} IRT emission line (FWHM $\leq 100 \, \mathrm{km \cdot s^{-1}}$).
}
  \label{fig:MdotFlux}
\end{center}
\end{figure}

$F^{\prime}_{\rm IRT}$ are listed in \tabref{tab:result_of_obs2}. The relation between the mass accretion rate, $\dot{M}$, and the surface flux of the Ca\,\emissiontype{II} line, $F^{\prime}_{\lambda 8662}$, is shown in \figref{fig:MdotFlux}. $\dot{M}$ listed in \tabref{tab:objects1} were taken from several studies, in which the rates were estimated from the amount of veiling: \citet{najita} and \citet{ca04} estimated $\dot{M}$ with a continuum component between the visible and UV wavelengths, \citet{gu98}, \citet{wg02}, \citet{h98}, and \citet{i13} determined $\dot{M}$ from the amount of the continuum veiling in the $U$-band. For objects whose veiling had not been determined, we referred $\dot{M}$ estimated with an EQW of the H$\mathrm{\alpha}$ emission line (\cite{l04}).

In \figref{fig:MdotFlux}, the circle symbols represent the PMS stars in the molecular clouds and the square symbols are the PMS stars in the moving groups. The open symbols show objects with a broad Ca\,\emissiontype{II} IRT emission line (FWHM $> 100 \, \mathrm{km \cdot s^{-1}}$) and the filled symbols represent the objects with a narrow Ca\,\emissiontype{II} IRT emission line (FWHM $\leq 100 \, \mathrm{km \cdot s^{-1}}$). As mentioned above, the broad line components are well explained by the magnetospheric accretion model in \citet{mu98}, while the narrow emission Ca\,\emissiontype{II} lines are generated in the stellar chromosphere (\cite{hp92}). 

We can see two groups among the PMS stars. High mass accretion rate PMS stars ($\dot{M} \gtrsim 10^{-7} \, \mathrm{M_{\odot} \cdot yr^{-1}}$) have a broad Ca\,\emissiontype{II} IRT emission line. DG Tau, DL Tau, and DR Tau belong to this group. Those are classified as accretors based on the accretion diagnosis of \citet{mo05}. The relationship between the mass accretion rate and the surface flux of the Ca\,\emissiontype{II} IRT emission line is consistent with that for the accretors (\cite{mo05}). The mass accretion rate does not show any correlation with $F^{\prime}_{\lambda 8662}$ for the objects with a narrow emission and almost all objects have $\log F^{\prime}_{\lambda 8662} \sim 3$ with a flat distribution to $\dot{M}$. Those do not meet the accretor criteria presented by \citet{mo05}. 
Most objects in our sample, including CTTSs, belong to this group. Any investigated PMS stars with a low mass accretion rate, except SU Aur and RECX 15, does not show broad Ca\,\emissiontype{II} IRT emission lines.
The difference of $F^{\prime}_{\rm IRT}$ between the two groups is more than 1 order of magnitude.
We found that there is a significant difference between the group of the objects with a broad Ca\,\emissiontype{II} IRT emission line and the group of objects with a narrow Ca\,\emissiontype{II} IRT emission line. This has been well established difference between accretors and non-accretors by some studies, and our result supports that. 

The Ca\,\emissiontype{II} IRT emission lines of PMS stars vary with time (e.g., \cite{jb97}).
We present the EQWs and profiles of the Ca\,\emissiontype{II} IRT emission lines of BP Tau in \tabref{tab:bptau}.
In this work, BP Tau shows narrow emissions superposed on broad absorption features.
The $W_{\lambda 8662}$ of BP Tau in this study is $1.15 \, \mathrm{\AA}$.
In contrast, the Ca\,\emissiontype{II} IRT lines in \citet{mo05} are broad and $W_{\lambda 8662}$ is $7.8 \, \mathrm{\AA}$.
\citet{mo05} referred the EQWs of the Ca\,\emissiontype{II} IRT emission lines in \citet{mu98}, BP Tau as observed in 1996, has strong $F^{\prime}_{\lambda 8662}$, similar to DG Tau, DL Tau and DR Tau. 
In \citet{mo05}, objects with broad Ca\,\emissiontype{II} IRT emission lines as classified by Muzerolle et al. (1998) show a positive correlation with the mass accretion rate.
Most objects in this study have narrow components, so that their $F^{\prime}_{\lambda 8662}$ has no correlation with the mass accretion rate.

\begin{tiny}

\begin{table}[tbp]
  \centering
  \caption{Time variation of the profiles of the CTTS, BP Tau}
  \begin{tabular}{llll} \hline
Observation & Profiles & $W_{\lambda 8662} \, \mathrm{[\AA]}$ & Reference \\
Date &  &  &  \\ \hline
1993 & Narrow emission ($\lambda 8498$) & Not measured & \citet{ar02}\\
1996 & Broad emission + Narrow emission ($\lambda 8542$) & 7.8 & \citet{mu98} \\
2006 & Broad emission + Narrow emission ($\lambda 8498, 8542, 8662$) & 6.44 & \citet{m12} \\
2008 & Narrow emission + Photospheric abrorption ($\lambda 8498, 8542, 8662$) & 1.15 & This work\\
 \hline
  \end{tabular}
\label{tab:bptau}
\end{table}

\end{tiny}

\section{Rotation-Activity Relation}
\label{thiswork : rotation-activity}

The narrow Ca\,\emissiontype{II} IRT emission lines in most of our samples are not correlated with the mass accretion rate.
\citet{noyes} used the Rossby number, $N_{\rm R}$, as an indicator of stellar dynamo activity.
We calculated $N_{\rm R}$ as follows, 
\begin{equation}
	N_{\rm R} = \frac{2 \pi R_*}{\tau_{\rm c} v \sin i},
\end{equation}
where $v \sin i$ is taken from Catalog of Stellar Rotational Velocities (\cite{g05}), \citet{n12}, \citet{to06}, and \citet{me11}.
For objects whose $v \sin i$ was not measured, we took the rotation period from Messina et al. (2010, 2011, 2018), \citet{h18}, and the AAVSO International Variable Star Index (\cite{wa06}). 
The rotation periods of IT Tau, TWA 06, TWA 14, TWA 23, and RECX 09 are 7.56 days, 0.54 days, 0.63 days, 1.03 days, and 1.71 days, respectively.
$R_*$ represents the stellar radius and $\tau_{\rm c}$ is the convective turnover time.

We estimate $\tau_{\rm c}$ of our PMS stars using pre-main sequence evolutionary tracks and convective turnover time of stars with $0.065-5.0 \, \mathrm{M_{\odot}}$ presented in \citet{jk07}.
The convective turnover time of a solar-mass PMS is between 300 and 100 days for the CTTS phase, and between a few hundred days and several dozens of days for the WTTS phase.
The convective turnover time of an object near and on ZAMS remains stable at a few dozen days.
In \citet{jk07}, the $\tau_{\rm c}$ of low mass stars near the main sequence was not calculated.
For objects whose $\tau_{\rm c}$ was not calculated in \citet{jk07}, we apply the approximation from \citet{noyes} when calculating $\tau_{\rm c}$ with $(B-V)_0$.
$(B-V)_0$ was calculated from $B-V$ (\cite{v07}, \cite{ma07}, \cite{ma06}, \cite{he16}, \cite{hb88}, \cite{ha03}, and \cite{d14}) and $A_V$ in \citet{k09}, \citet{ma06}, and \citet{wa10}.

We calculated the ratio of the surface flux of the Ca\,\emissiontype{II} IRT lines to the stellar bolometric luminosity, $R^{\prime}_{\rm IRT}$, for each Ca\,\emissiontype{II} IRT line.
$R^{\prime}_{\rm IRT}$ is similar to the parameter $R^{\prime}_{\rm HK}$ derived from the Ca\,\emissiontype{II} H and K lines, as described by \citet{noyes}. 
$R^{\prime}_{\rm \lambda 8542}$, the ratio of the surface flux of the $\lambda 8542 \, \mathrm{\AA}$ line to the stellar bolometric luminosity, has been previously used by \citet{so93} and \citet{jj97}.
In addition to $R^{\prime}_{\rm \lambda 8542}$, $R^{\prime}_{\lambda 8498}$ and $R^{\prime}_{\rm \lambda 8662}$ were used by \citet{marsden09}.
In their study, $R^{\prime}_{\rm \lambda 8498}$ and $R^{\prime}_{\rm \lambda 8662}$ show qualitatively similar results to $R^{\prime}_{\rm \lambda 8542}$.
$R^{\prime}_{\rm IRT}$ involves $R^{\prime}_{\rm \lambda 8498}$, $R^{\prime}_{\rm \lambda 8542}$ and $R^{\prime}_{\rm \lambda 8662}$, which are determined individually for each of the Ca\,\emissiontype{II} IRT lines.
For calculating $R^{\prime}_{\rm IRT}$, $F^{\prime}_{\rm IRT}$ are divided by $\sigma T_{\rm eff}^4$.
\begin{equation}
	R^{\prime}_{\rm IRT} = \frac{F^{\prime}_{\rm IRT}}{\sigma T_{\rm eff}^4},
\end{equation}
where $\sigma$ is Stefan-Boltzmann's constant.
The dependence of the surface flux upon the $T_{\rm eff}$ of the objects is eliminated by this calculation.
$R^{\prime}_{\rm IRT}$ are listed in \tabref{tab:result_of_obs2}.

\begin{small}

\renewcommand{\tabcolsep}{4pt}  
\begin{longtable}{p{36mm}p{12mm}p{12mm}p{12mm}p{12.5mm}p{12.5mm}p{12.5mm}p{12mm}p{10mm}}
\caption{$F^{\prime}_{\rm IRT}$, and $R^{\prime}_{\rm IRT}$ of the Ca\emissiontype{II} IRT emission lines ($\lambda$ 8498, 8542, 8662 $A$)}
\label{tab:result_of_obs2}
\hline
Object & $\log F^{\prime}_{\lambda 8498}$ &  $\log F^{\prime}_{\lambda 8542}$ &  $\log F^{\prime}_{\lambda 8662}$ &  $\log R^{\prime}_{\lambda 8498}$ &  $\log R^{\prime}_{\lambda 8542}$ &  $\log R^{\prime}_{\lambda 8662}$ &  $\log N_{\rm R}$  & note\\

{}    & $\mathrm{W \cdot m^{-2}}$  & $\mathrm{W \cdot m^{-2}}$ &    $\mathrm{W \cdot m^{-2}}$     &             &  &          &      &    \\
{} &   (1)  &  (2) &  (3) &  (4) &  (5) &  (6) & (7) & (8)\\
\hline
\endhead
\endfoot
\multicolumn{2}{@{}l@{}}{\hbox to0pt{\parbox{170mm}{ 
{(8) This column shows the groups based on the FWHM of the Ca II emission lines ($\lambda 8498 \, \mathrm{\AA}$) and the membership of the objects. NC: Narrow emission objects belonging to the molecular clouds, BC: Broad emission objects belonging to the molecular clouds, NM: Narrow emission objects belonging to the moving groups, BM: Broad emission objects belonging to the moving groups} }\hss}} 
\endlastfoot
\multicolumn{2}{l}{Taurus-Auriga molecular cloud}   &  &         &             &  &          &     &    \\ \hline
AA Tau                  &                         2.58 &                         2.65 &                         2.55 &                        -4.61 &                        -4.54 &                        -4.63 &    -1.71 &   NC \\
BP Tau                  &                         3.50 &                         3.70 &                         3.54 &                        -3.80 &                        -3.59 &                        -3.76 &    -1.57 &   NC \\
CX Tau                  &                         2.57 &                         2.69 &                         2.66 &                        -4.32 &                        -4.20 &                        -4.23 &    -1.75 &   NC \\
CoKu Tau4               &                         3.38 &                         3.57 &                         3.43 &                        -3.81 &                        -3.62 &                        -3.76 &    -2.00 &   NC \\
DF Tau                  &                         2.99 &                         3.24 &                         3.07 &                        -4.02 &                        -3.77 &                        -3.94 &    -1.53 &   NC \\
DG Tau                  &                         4.79 &                         4.77 &                         4.81 &                        -2.25 &                        -2.27 &                        -2.23 &    -1.85 &   BC \\
DL Tau                  &                         5.15 &                         5.12 &                         5.01 &                        -2.01 &                        -2.04 &                        -2.15 &    -1.83 &   BC \\
DM Tau                  &                         3.08 &                         3.19 &                         3.11 &                        -3.98 &                        -3.87 &                        -3.95 &    -1.71 &   NC \\
DR Tau                  &                         5.90 &                         5.95 &                         5.86 &                        -1.40 &                        -1.35 &                        -1.44 &    -1.81 &   BC \\
DS Tau                  &                         3.24 &                         3.44 &                         3.38 &                        -3.94 &                        -3.74 &                        -3.79 &    -1.68 &   NC \\
FN Tau                  &                         3.78 &                           -  &                         3.83 &                        -3.49 &                           -  &                        -3.44 &    -1.17 &   NC \\
FP Tau                  &                         2.49 &                           -  &                         2.48 &                        -4.44 &                           -  &                        -4.44 &    -1.97 &   NC \\
GH Tau                  &                         2.41 &                         2.47 &                         2.36 &                        -4.56 &                        -4.50 &                        -4.61 &    -1.68 &   NC \\
GM Aur                  &                         3.26 &                         3.54 &                         3.45 &                        -4.04 &                        -3.76 &                        -3.85 &    -1.55 &   NC \\
GO Tau                  &                         3.15 &                         3.29 &                         3.21 &                        -4.00 &                        -3.86 &                        -3.94 &    -1.57 &   NC \\
HBC 374                 &                         2.85 &                         3.08 &                           -  &                        -4.32 &                        -4.08 &                           -  &    -1.67 &   NC \\
HBC 376                 &                         3.50 &                         3.63 &                           -  &                        -3.82 &                        -3.69 &                           -  &       -  &   NC \\
HBC 407                 &                         2.82 &                         2.91 &                         2.83 &                        -4.61 &                        -4.52 &                        -4.60 &    -1.03 &   NC \\
HBC 427                 &                         2.83 &                         2.95 &                         2.90 &                        -4.44 &                        -4.32 &                        -4.37 &    -1.45 &   NC \\
HD 285778               &                         3.06 &                         3.27 &                           -  &                        -4.61 &                        -4.40 &                           -  &    -1.33 &   NC \\
HP Tau                  &                           -  &                           -  &                           -  &                           -  &                           -  &                           -  &    -2.54 &   NC \\
IT Tau                  &                         3.57 &                         3.67 &                         3.60 &                        -3.56 &                        -3.45 &                        -3.52 &    -1.33 &   NC \\
LkCa 04                 &                         2.69 &                         2.86 &                         2.82 &                        -4.30 &                        -4.13 &                        -4.17 &    -1.87 &   NC \\
LkCa 14                 &                         2.88 &                         3.02 &                         2.94 &                        -4.37 &                        -4.24 &                        -4.31 &    -1.90 &   NC \\
LkCa 15                 &                         3.13 &                         3.15 &                         3.13 &                        -4.11 &                        -4.10 &                        -4.12 &    -1.66 &   NC \\
LkCa 19                 &                         3.13 &                         3.27 &                         3.14 &                        -4.34 &                        -4.21 &                        -4.34 &    -1.68 &   NC \\
RY Tau                  &                         4.38 &                         4.26 &                         4.19 &                        -3.41 &                        -3.53 &                        -3.60 &    -1.19 &   BC \\
SU Aur                  &                         3.49 &                         3.72 &                         3.63 &                        -3.82 &                        -3.59 &                        -3.68 &    -1.92 &   BC \\
UX Tau                  &                         3.26 &                         3.39 &                         3.33 &                        -4.08 &                        -3.95 &                        -4.01 &    -1.52 &   NC \\
V1023 Tau               &                         3.30 &                         3.47 &                           -  &                        -3.87 &                        -3.69 &                           -  &    -1.67 &   NC \\
V1204 Tau               &                         3.08 &                         3.11 &                           -  &                        -4.40 &                        -4.37 &                           -  &    -1.54 &   NC \\
V1297 Tau               &                         3.04 &                         3.11 &                         3.02 &                        -4.52 &                        -4.45 &                        -4.55 &    -1.01 &   NC \\
V1321 Tau               &                         2.99 &                         3.12 &                         3.04 &                        -4.19 &                        -4.06 &                        -4.13 &    -1.28 &   NC \\
V1348 Tau               &                         2.95 &                         3.08 &                         2.98 &                        -4.39 &                        -4.26 &                        -4.37 &    -1.09 &   NC \\
V830 Tau                &                         3.00 &                         3.20 &                         3.03 &                        -4.17 &                        -3.97 &                        -4.14 &    -1.98 &   NC \\
V836 Tau                &                         2.81 &                         2.90 &                         2.88 &                        -4.19 &                        -4.09 &                        -4.11 &    -1.58 &   NC \\
ZZ Tau                  &                         3.24 &                         3.37 &                         3.20 &                        -3.98 &                        -3.85 &                        -4.02 &    -1.65 &   NC \\ \hline

\multicolumn{2}{l}{Orionis OB 1c association}   &  &         &             &  &          &      &   \\ \hline
HBC 167                 &                         3.96 &                         4.19 &                         4.06 &                        -3.75 &                        -3.52 &                        -3.65 &    -1.26 &   NC \\ \hline

\multicolumn{2}{l}{Upper Scorpius association}   &  &         &             &  &          &      &   \\ \hline
1RXS J161951.4-215431   &                          - &                          - &                          - &                          - &                          - &                          - &      - & NC \\ \hline

\multicolumn{2}{l}{Perseus molecular cloud}   &  &         &             &  &          &   &      \\ \hline
LkH$\rm \alpha$ 86      &                         2.65 &                         2.79 &                         2.69 &                        -4.36 &                        -4.22 &                        -4.32 &    -1.54 &   NC \\
LRL 72 &                           -  &                           -  &                           -  &                           -  &                           -  &                           -  &    -1.32 &   NC \\ \hline

\multicolumn{2}{l}{AB Doradus moving group}    &  &         &             &  &          &     &    \\ \hline
HIP 17695               &                         2.21 &                         2.36 &                         2.21 &                        -4.64 &                        -4.49 &                        -4.64 &    -2.15 &   NM \\ \hline

\multicolumn{2}{l}{$\rm \beta$ Pictoris moving group}   &  &         &             &  &          &      &   \\ \hline
HD 197481               &                         2.56 &                         2.64 &                         2.56 &                        -4.44 &                        -4.36 &                        -4.45 &    -1.70 &   NM  \\ \hline

\multicolumn{2}{l}{$\rm \eta$ Chamaeleontis cluster}   &  &         &             &  &          &    &     \\ \hline
RECX 04                 &                         2.83 &                         2.97 &                         2.85 &                        -4.34 &                        -4.20 &                        -4.33 &    -1.30 &   NM \\
RECX 06                 &                         2.50 &                         2.58 &                         2.44 &                        -4.44 &                        -4.36 &                        -4.51 &    -2.04 &   NM \\
RECX 07                 &                         2.94 &                         3.10 &                         3.00 &                        -4.36 &                        -4.20 &                        -4.29 &    -2.01 &   NM \\
RECX 09                 &                         2.48 &                         2.75 &                         2.60 &                        -4.66 &                        -4.39 &                        -4.53 &    -1.79 &   NM \\
RECX 10                 &                         2.80 &                         2.91 &                         2.79 &                        -4.40 &                        -4.29 &                        -4.41 &    -1.22 &   NM \\
RECX 11                 &                         3.06 &                         3.23 &                         3.15 &                        -4.27 &                        -4.10 &                        -4.18 &    -1.50 &   NM \\
RECX 15                 &                         3.85 &                         3.85 &                         3.74 &                        -3.19 &                        -3.18 &                        -3.29 &    -2.11 &   BM \\ \hline

\multicolumn{2}{l}{''Cha-Near'' region}   &  &         &             &  &          &      &   \\ \hline
RX J1147.7-7842         &                         2.85 &                         3.07 &                         2.90 &                        -4.26 &                        -4.04 &                        -4.21 &       -  &   BM \\
RX J1204.6-7731         &                         2.62 &                         2.76 &                         2.62 &                        -4.35 &                        -4.22 &                        -4.35 &    -1.44 &   NM \\ \hline

\multicolumn{2}{l}{TW Hydrae association}   &  &         &             &  &          &    &     \\ \hline
TWA 01                  &                         3.23 &                         3.51 &                         3.45 &                        -4.03 &                        -3.75 &                        -3.81 &    -1.53 &   NM \\
TWA 06                  &                         3.03 &                         3.24 &                         3.14 &                        -4.24 &                        -4.04 &                        -4.13 &    -2.21 &   NM \\
TWA 07                  &                         2.89 &                         2.96 &                         2.84 &                        -4.28 &                        -4.21 &                        -4.33 &    -1.20 &   NM \\
TWA 14                  &                         2.87 &                         3.05 &                         2.92 &                        -4.23 &                        -4.05 &                        -4.18 &    -2.57 &   NM \\
TWA 22                  &                         1.99 &                         2.27 &                         2.15 &                        -4.58 &                        -4.30 &                        -4.42 &    -1.05 &   NM \\
TWA 23                  &                         2.34 &                         2.46 &                         2.24 &                        -4.60 &                        -4.47 &                        -4.69 &    -2.34 &   NM \\
TWA 25                  &                         2.95 &                         3.07 &                         3.01 &                        -4.22 &                        -4.10 &                        -4.16 &    -1.36 &   NM \\ \hline \\
\end{longtable}
\renewcommand{\tabcolsep}{6pt}
\end{small}

The chromospheric activity of ZAMS stars is considered to be induced by dynamo activity.
\citet{marsden09} investigated the Ca\,\emissiontype{II} IRT emission lines of low-mass stars in the young open clusters, IC 2391 and IC 2602. 
For stars with $\log N_{\rm R} \leq -1.1$, $R^{\prime}_{\rm IRT}$ is constant at levels of  $\log R^{\prime}_{\rm \lambda 8498} \sim -4.4$, $\log R^{\prime}_{\rm \lambda 8542} \sim -4.2$, and $\log R^{\prime}_{\rm \lambda 8662} \sim -4.3$.
These regions are called the saturated regime.
For stars with $\log N_{\rm R} \geq -1.1$, $R^{\prime}_{\rm IRT}$ decreases with increasing $N_{\rm R}$.
This region is called the unsaturated regime.
\citet{marsden09} suggest that the chromosphere is completely filled by the emitting region for stars in the saturated regime.

\begin{figure}[h]
 \begin{minipage}[b]{1.0\linewidth}
  \centering
  \vspace{-0.5cm}
  \includegraphics[keepaspectratio, scale=0.35]{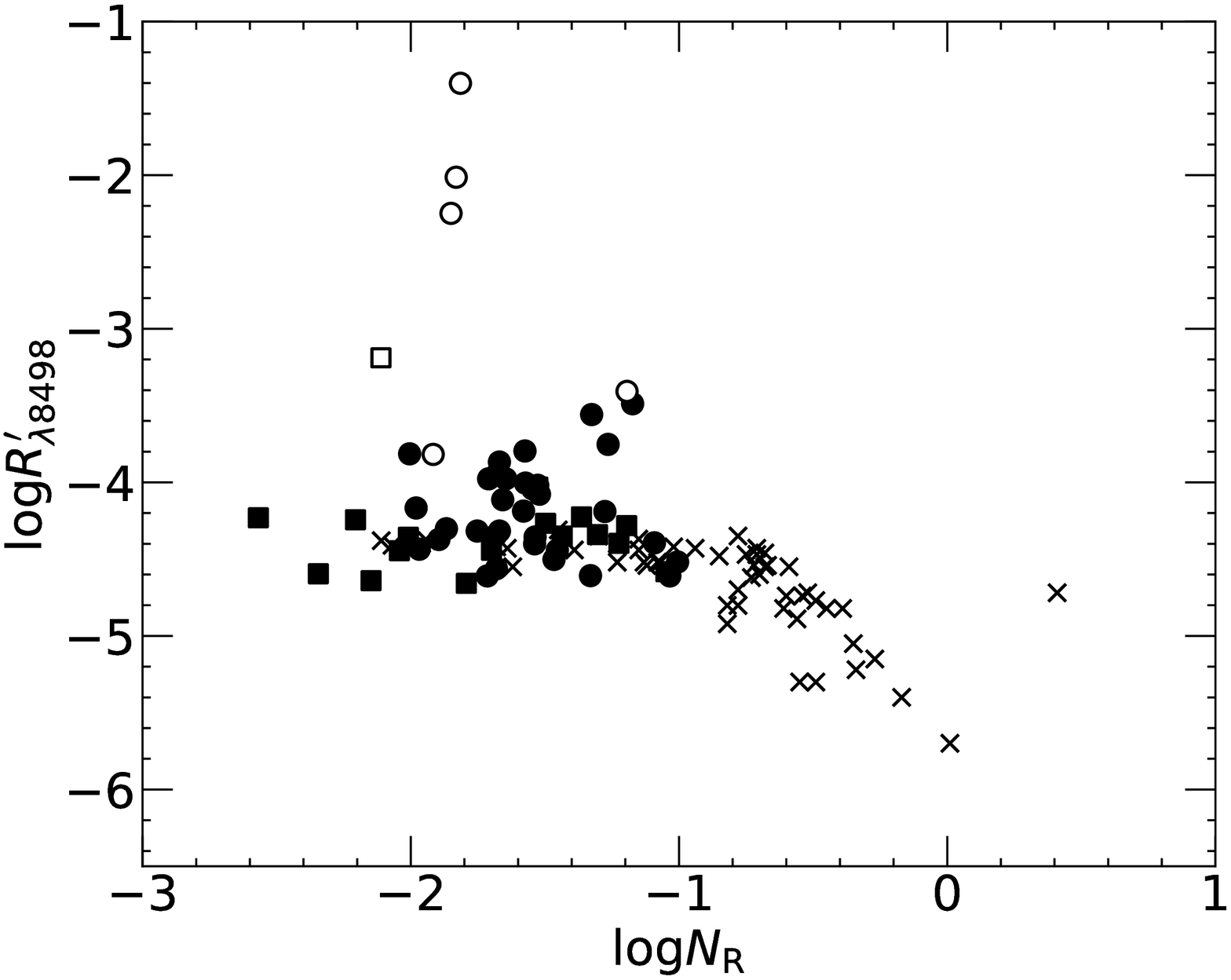}
  \vspace{-0.3cm}
  \subcaption{CaI\hspace{-.1em}I $\lambda 8498 \, \mathrm{\AA}$}
\label{fig:JKnomi1}
 \end{minipage}\\

 \begin{minipage}[b]{1.0\linewidth}
  \centering
  \vspace{-0.5cm}
  \includegraphics[keepaspectratio, scale=0.35]{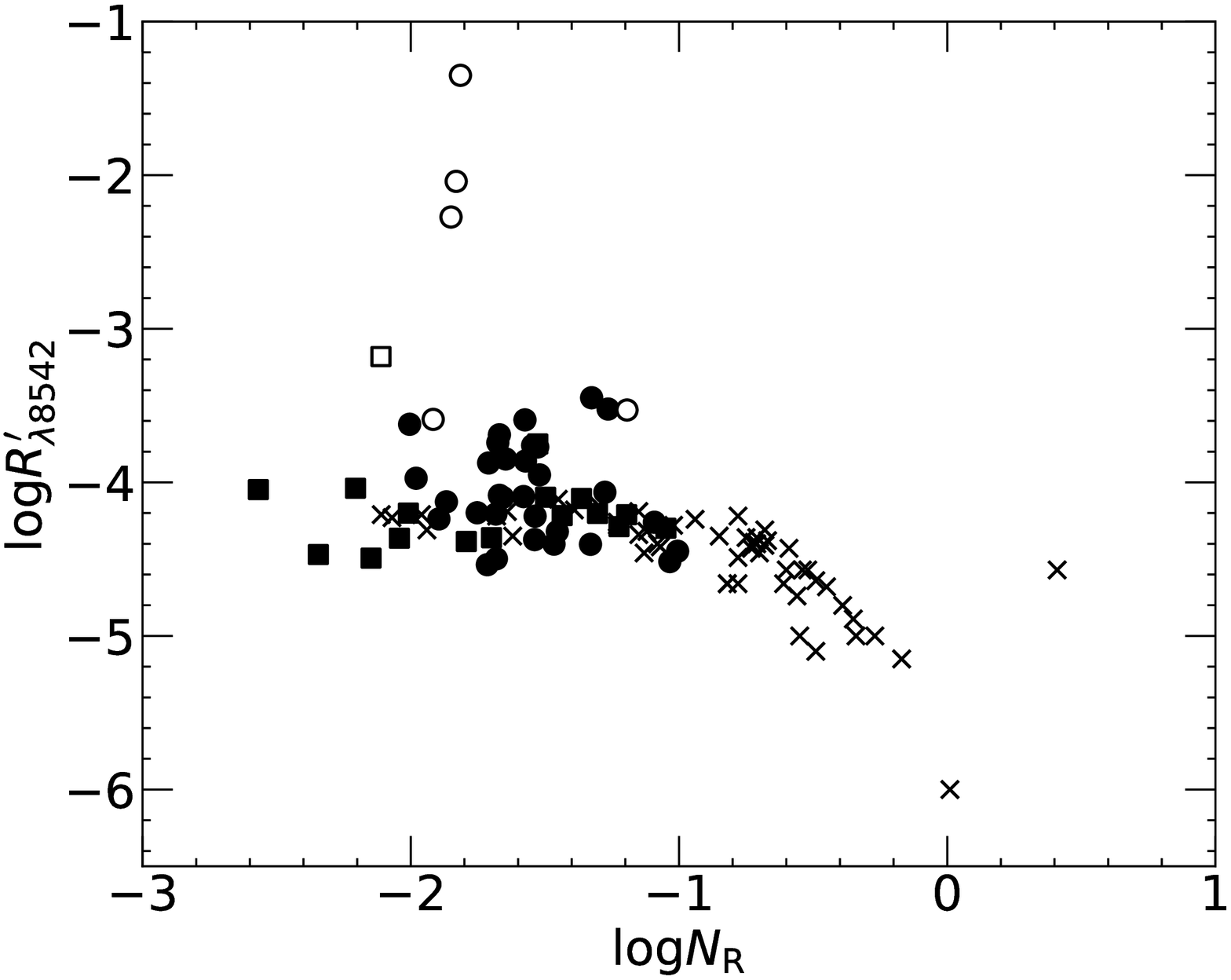}
  \vspace{-0.3cm}
  \subcaption{CaI\hspace{-.1em}I $\lambda 8542 \, \mathrm{\AA}$}
\label{fig:JKnomi2}
 \end{minipage}\\

 \begin{minipage}[b]{1.0\linewidth}
  \centering
  \vspace{-0.5cm}
  \includegraphics[keepaspectratio, scale=0.35]{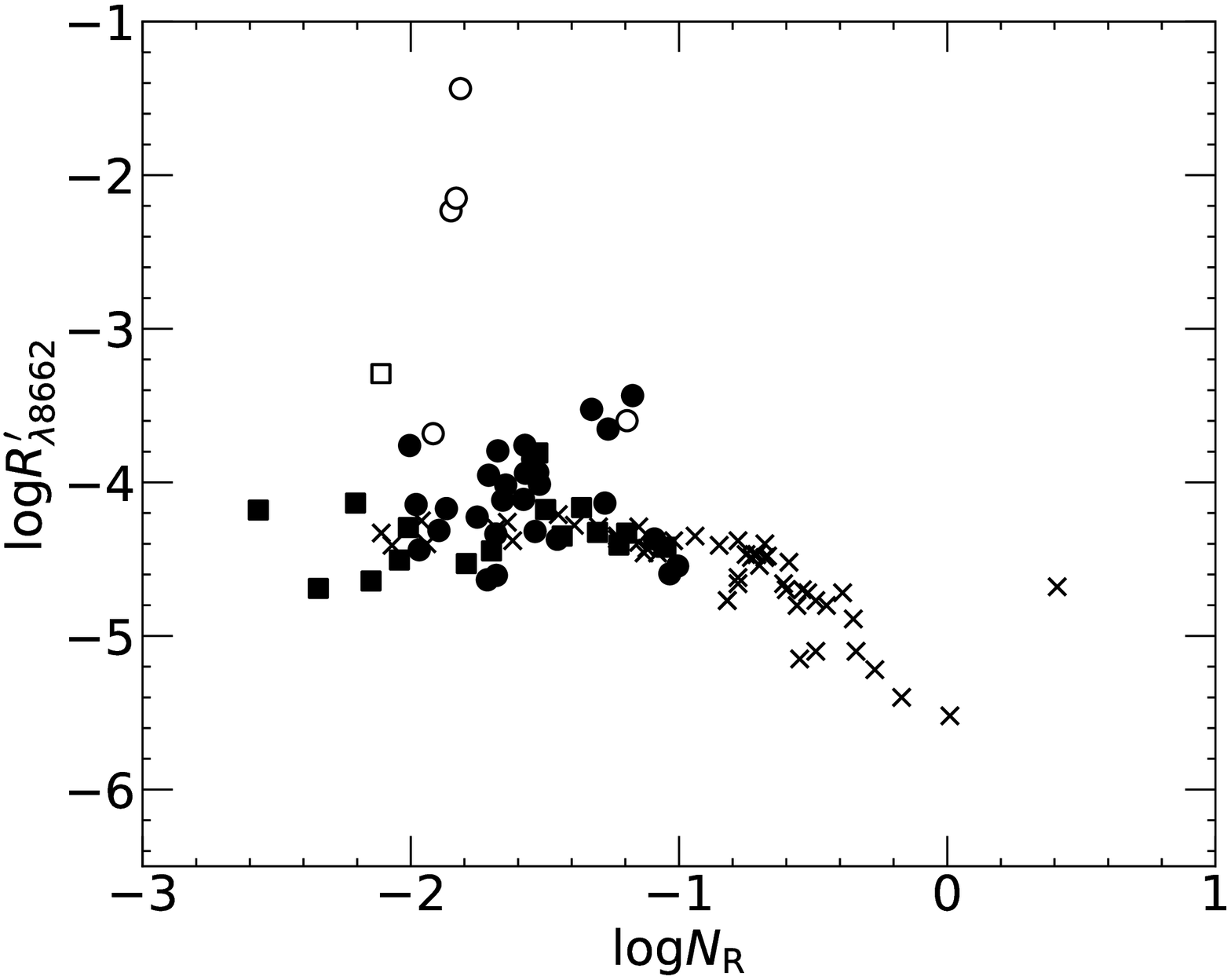}
  \vspace{-0.3cm}
  \subcaption{CaI\hspace{-.1em}I $\lambda 8662 \, \mathrm{\AA}$}
\label{fig:JKnomi3}
 \end{minipage}
 \caption{Relation between the ratio of the surface flux of the Ca\,\emissiontype{II} IRT line to the stellar bolometric luminosity, $R^{\prime}_{\rm IRT}$ and the Rossby number, $N_{\rm R}$.
The circle symbols represent PMS stars in molecular clouds and the square symbols indicate PMS stars in moving groups.
The filled symbols represent PMS stars with narrow Ca\,\emissiontype{II} IRT emission lines and the open symbols show PMS stars with broad Ca\,\emissiontype{II} IRT emission lines.
The cross symbols represent low-mass stars in the young open clusters IC 2391 and IC 2602, as studied in \citet{marsden09}.}
\label{fig:JKnomi}
\end{figure}

\figref{fig:JKnomi} shows $R^{\prime}_{\rm IRT}$ as a function of $N_{\rm R}$.
The circle symbols represent PMS stars in molecular clouds and the square symbols are PMS stars in moving groups.
The open circles and squares show objects with a broad Ca\,\emissiontype{II} IRT emission line and the filled circles and squares represent objects with a narrow Ca\,\emissiontype{II} IRT emission line.
The cross symbols represent low-mass stars in IC 2391 and IC 2602, as studied in \citet{marsden09}.

Most PMS stars have $N_{\rm R}$ and $R^{\prime}_{\rm IRT}$, which is similar to those of the cluster members in the saturated regime. We claim that the chromospheres of the PMS stars are activated by the magnetic field generated by the dynamo process and that the chromosphere of these stars are completely filled by the emitting region. It is consistent with the previous studies indicating that PMS stars have optically thick Ca\,\emissiontype{II} IRT emission lines (e. g. \cite{bb93}; \cite{hp92}) and our result supports them.
In contrast, three PMS stars with strong and broad emission lines (DG Tau, DL Tau and DR Tau) have a $R^{\prime}_{\rm IRT}$ 2 orders of magnitude larger than the cluster members. We consider that strong emission lines are caused by heavy mass accretion from their protoplanetary disks.

Several researchers have pointed out that low-mass dwarfs showing strong Ca\,\emissiontype{II} emission lines also have strong X-ray emissions (\cite{ho17}), strong magnetic fields (\cite{l17} ; \cite{vj01}), and that the flux variation of the Ca\,\emissiontype{II} lines is related to the coverage of their faculae or spots. Some of these features have already been found in low-mass PMS stars. Furthermore, it has to be investigated whether low-mass PMS stars have these other features.
\clearpage
\chapter{Conclusion}
\label{colclusion}


In this paper, we investigated the Ca\,\emissiontype{II} infrared triplet lines ($\lambda 8498 \cdot 8542 \cdot 8662 \, \mathrm{\AA}$) of 60 PMS stars.
The observations were conducted with Nayuta/MALLS and Subaru/HDS.
Archived data obtained from the Keck/HIRES, VLT/UVES, and VLT/X-Shooter were also used.

\begin{enumerate}


  \item The ratios of the equivalent widths of the Ca\,\emissiontype{II} IRT emission lines $W_{\lambda 8542} / W_{\lambda 8498}$ are 1.0--2.0 and $W_{\lambda 8662}/W_{\lambda 8498}$ are 0.7--1.7 for both PMS stars and the low-mass stars in the young open clusters (Marsden, Carter \& Donati 2009).
This suggests that the Ca\,\emissiontype{II} IRT emission lines originate from the regions analogous to solar plages and flares.


  \item Seven PMS stars (DG Tau, DL Tau, DR Tau, RY Tau, SU Aur, RECX 15, and RX J1147.7-7842) have broad Ca\,\emissiontype{II} IRT emission lines. 
It is suggested that their broad emissions result from heavy mass accretion from their protoplanetary disks.


  \item Most PMS stars have narrow Ca\,\emissiontype{II} IRT emission lines similar to low-mass stars in young open clusters.
The emissions of these objects indicate no correlation with the mass accretion rate.
The surface flux of the Ca\,\emissiontype{II} IRT emission lines to the stellar bolometric luminosity, $R^{\prime}_{\rm IRT}$, of these objects are large, as the largest $R^{\prime}_{\rm IRT}$ of the cluster members.
Most PMS stars have chromospheric activity similar to zero-age main-sequence stars.
The chromosphere of these stars are completely filled by the Ca\,\emissiontype{II} emitting region.

\end{enumerate}


\begin{ack}
This research has made use of the Keck Observatory Archive (KOA), which is operated by the W. M. Keck Observatory and NASA Exoplanet Science Institute (NExScI), and it is under contract with the National Aeronautics and Space Administration, and is based on the observations made with ESO Telescopes at the La Silla Paranal Observatory under programmes ID 075.C-0321, 082.C-0005, 084.C-1095, 085.C-0238, 086.C-0173, 094.C-0327, 094.C-0805, and 094.C-0913.
\end{ack}

\begin{quote}
  
\end{quote}

\end{document}